Monovalent ions modulate the flux through multiple folding pathways of an RNA pseudoknot

Jorjethe Roca[1#], Naoto Hori[2], Yogambigai Velmurugu[1], Ranjani Narayanan[1‡], Prasanth Narayanan[1], D. Thirumalai[2], and Anjum Ansari[1,3*]

[1]Department of Physics and [3]Department of Bioengineering,

University of Illinois at Chicago, Chicago, IL 60607

[2]Department of Chemistry, University of Texas at Austin, Austin, TX 78712

[#]Present address: Thomas C. Jenkins Department of Biophysics, Johns Hopkins University,

Baltimore MD 21218

[‡]Present address: Division of Physics & Applied Physics, School of Physical and Mathematical

Sciences, Nanyang Technological University, Singapore 637371

[*]Corresponding author.

E-mail: ansari@uic.edu. Phone: (312) 996–8735. FAX: (312) 996–9016

**ABSTRACT**

The functions of RNA pseudoknots (PKs), which are minimal tertiary structural motifs and an integral part of several ribozymes and ribonucleoprotein complexes, are determined by their structure, stability and dynamics. Therefore, it is important to elucidate the general principles governing their thermodynamics/folding mechanisms. Here, we combine experiments and simulations to examine the folding/unfolding pathways of the VPK pseudoknot, a variant of the Mouse Mammary Tumor Virus (MMTV) PK involved in ribosomal frameshifting. Fluorescent nucleotide analogs (2-aminopurine and pyrrolocytidine) placed at different stem/loop positions in the PK, and laser temperature-jump approaches serve as local probes allowing us to monitor the order of assembly of VPK with two helices with different intrinsic stabilities. The experiments and molecular simulations show that at 50 mM KCl the dominant folding pathway populates only the more stable partially folded hairpin. As the salt concentration is increased a parallel folding pathway emerges, involving the less stable hairpin structure as an alternate intermediate. Notably, the flux between the pathways is modulated by the ionic strength. The findings support the principle that the order of PK structure formation is determined by the relative stabilities of the hairpins, which can be altered by sequence variations or salt concentrations. Our study not only unambiguously demonstrates that PK folds by parallel pathways, but also establishes that quantitative description of RNA self-assembly requires a synergistic combination of experiments and simulations.

## SIGNIFICANCE

The assembly mechanism of RNA, vital to describing their functions, depends on both the sequence and the metal ion concentration. How the latter influences the folding trajectories remains an important unsolved problem. Here, we examine the folding pathways of an RNA pseudoknot (PK) with key functional roles in transcription and translation, using a combination of experiments and simulations. We demonstrate that the PK, consisting of two helices with differing stabilities, folds by parallel pathways. Surprisingly, the flux between them is modulated by monovalent salt concentration. Our work shows that the order of assembly of PKs is determined by the relative stability of the helices, implying that the folding landscape can be controlled by sequence and ion concentration.

\body

## INTRODUCTION

The remarkable discovery that RNA can function as an enzyme and subsequent discoveries showing that RNA molecules are involved in many cellular functions have elevated the need to describe how they fold (1). The functions of RNA molecules are determined by their three-dimensional folded structures and on the underlying folding landscapes with potentially many minima separated by free energy barriers, which implies that their dynamics is governed by transitions between the different energy states (2). Therefore, elucidating the folding mechanisms of RNA molecules and the timescales in which they interconvert between different states is critical for understanding how structure and dynamics influence their functions. Such studies can be leveraged into the functional characterization of newly discovered RNAs, and could lead to designer RNA molecules with desirable functions.

It has been firmly established that RNA self-assembly occurs through multiple pathways, described by the kinetic partitioning mechanism (KPM), according to which a fraction of molecules fold rapidly and another fraction fold slowly as they become trapped in misfolded conformations corresponding to the minima in the rugged folding landscape (3). Evidence for the KPM was first obtained for the *Tetrahymena* ribozyme (4). Subsequently it has also been determined that the folding landscapes of even RNA hairpins are rugged (5). The relatively fast folding times of nucleic acid hairpins and the ability to enumerate all possible folded and misfolded conformations have made hairpins particularly useful as model systems to validate various computational studies of the folding mechanisms (6–14). However, hairpins lack tertiary interactions, and therefore they are limited in providing insights into the assembly of larger RNA molecules.

Here, we have focused on the folding mechanisms of RNA pseudoknots (PKs) that serve as minimal tertiary structural motifs. In contrast to ribozymes that fold in times at the order of 1-1000 seconds (15, 16), we previously reported that PKs fold in a few milliseconds (17) in accord with predictions using simulations of coarse-grained RNA models (18). Thus, RNA pseudoknots provide a bridge between rapidly folding secondary structure elements and the slower reorganization of ribozymes. They are ideal systems to explore the interplay between competing secondary and tertiary interactions that determine the diverse folding pathways, and provide a way to investigate the role ions play in modulating the flux between different assembly pathways.

The PKs form when the nucleotides in a hairpin loop form base pairs with nucleotides downstream of that hairpin, creating a structure with two stems and two loops (Figure 1) with or without intervening elements (19). They are involved in several biological functions, related to replication, RNA processing and transcriptional and translational regulation (20–23). They also stabilize the structures of larger RNA assemblies such as the ribosome, ribozymes, and ribonucleoproteins (24, 25), and are often the rate-limiting step in their folding pathways. Thus, understanding the global principles that determine the thermodynamics and folding pathways of PKs could provide insights into their functions and into the self-assembly of larger RNA complexes.

Single-molecule pulling experiments and more recently, nanopore-induced unfolding experiments have shown that PK folding/unfolding can occur through several complex pathways that include the formation of transient hairpins and nonnative structures (26–34). These experiments have also unveiled that in many instances the RNA molecules do not fold into a pseudoknot folding conformation but instead fold into alternate structures (30). However, the energy landscapes and folding pathways inferred from pulling experiments could differ from those obtained by changing the temperature, as demonstrated for RNA hairpins (10, 35–37). NMR dynamics and single molecule FRET experiments have also been performed in the absence of force, but these have been restricted to studying the dynamics of interconversion between a pseudoknot and its intermediate folds (usually structures with folded hairpins), and thus have not provided a full picture of the folding/unfolding process of PKs (38–40).

Computational studies have predicted that PKs fold through diverse mechanisms depending on their sequence, either by parallel pathways or cooperatively in an all-or-none manner or by a combination of both mechanisms, exploring several intermediate states, some of which contain non-native base pairing or alternate structures (18, 41–46). The principle that assembly routes of PKs depends on the stabilities of individual secondary structural elements (18) explains the diversity of the folding mechanisms of PK. However, to our knowledge, experimental demonstration of folding through parallel pathways and the impact of ions in affecting the diversity of states is lacking. This is due in part to experimental limitations that do not allow probing of folding kinetics in the submillisecond region, where hairpins and possible folding intermediates could form.

Here, we use a laser temperature jump (T-jump) approach to study pseudoknot folding with submilli- and submicrosecond resolution by taking advantage of fluorescent nucleotide analogs (47–50) judiciously placed at various positions either in the stem or loop regions of a pseudoknot. We apply these fluorescent approaches to study the folding mechanisms of VPK, a variant of the MMTV PK that promotes efficient frameshifting in mouse mammary tumor virus (51–53). In order to ascertain how ions alter the flux between distinct folding pathways, we have used experiments and simulations to determine the thermodynamics and folding kinetics of VPK over a broad range of ionic conditions. The results provide an unparalleled experimental measure, confirmed by coarse-grained simulations, of the partitioning of the folding process into different folding pathways. We find that at 50 mM KCl, the dominant folding pathway of VPK populates the more stable PK's constituent hairpins as an intermediate. However, as the salt concentration is increased a parallel folding pathway emerges, involving the other less stable hairpin as an intermediate, with the flux through the alternative pathway increasing with increasing salt concentrations. The combined experiments and simulations directly demonstrate that pseudoknots can fold via parallel pathways. We further establish that monovalent ions can modulate the flux through distinct folding pathways. As enunciated previously (18), the order of assembly is largely determined by the relative stabilities of the hairpin, which can be changed by altering the ionic strength.

## RESULTS

**Fluorescent nucleotide analogs as probes of PK unfolding thermodynamics:** Previous studies on the thermodynamics and kinetics of VPK unfolding were carried out using either UV absorbance measurements (17, 54) or differential scanning calorimetry studies (54). While these studies demonstrated hierarchical unfolding of VPK, they could not discern whether unfolding occurred via parallel pathways. To probe experimentally VPK's unfolding pathways, here we used fluorescently-labeled VPK. We prepared two pseudoknot constructs in which we substituted one nucleotide of the VPK sequence for a fluorescent analog: VPK-2AP, with 2-aminopurine (2AP) instead of adenine at position 20, and VPK-pC, with pyrrolocytidine (pC) instead of cytosine at position 31 (Figure 1). We placed 2AP at a position adjacent to the end of stem 1 since VPK does not have a suitably placed adenine in either of its stems (other than at the 3'-end, which could be prone to "fraying" ends). The 2AP in this position is particularly sensitive to the conformational changes (see below) that occur along the unfolding pathway that involves the partially folded intermediate state pkHP1 (Figure 2a; top pathway). On the other hand, the pC location was chosen to directly monitor the loss of stem 2 either when the PK unfolds via pkHP1 (Figure 2d; top pathway) or when the residual pkHP2 unfolds (Figure 2d; bottom pathway).

To assess the degree to which the incorporated fluorescent nucleotides perturb the RNA structures, we performed absorbance melting measurements on unlabeled and labeled VPK (SI Figure S1). At 50 mM KCl, the absorbance melting profiles for all the VPK constructs show two transitions that were analyzed in terms of two sequential van't Hoff steps (see SI Methods 1.3). The thermodynamic parameters from this analysis are summarized in SI Table S1. The melting temperatures for the two transitions in the unlabeled VPK are at $T_{m1} \approx 47$ °C and $T_{m2} \approx 85$ °C, consistent with previous studies (17, 54). While 2AP has a negligible effect on the stability of the fully folded pseudoknot, with $T_{m1} \approx 49$ °C, it appears to significantly stabilize the partially unfolded intermediate hairpin states, with $T_{m2}$ shifted to ~ 95 °C. Our results for VPK-pC show that pC slightly stabilizes both the folded pseudoknot and the intermediate hairpin states, with $T_{m1} \approx 53$ °C and $T_{m2} \approx 89$ °C. Although incorporation of nucleotide analogs does have a detectable effect on the stability of the RNA structures, they nonetheless show promise in reporting on the nature of the transitions and the pathways for pseudoknot unfolding, as discussed below.

**Melting profiles reveal that VPK unfolds predominantly via unfolding of stem 2:** Next, we examined the melting profiles of VPK-2AP and VPK-pC from fluorescence measurements in the same ionic conditions as before (Figure 2). As in the absorbance studies, fluorescence studies on VPK-2AP also revealed two unfolding steps, with a decrease in 2AP fluorescence observed in the first transition followed by an increase in the second transition (Figure 2b). A decrease in fluorescence in the first transition is attributed to the loss of stem 2 in the PK structure, which likely enables the 2AP label to stack better over stem 1. These results suggest that unfolding via a pkHP1 intermediate (Figure 2a; top pathway) is the dominant trajectory in VPK unfolding; unfolding via pkHP2 where stem 1 melts first (Figure 2a; bottom pathway) is not expected to dramatically affect 2AP fluorescence. Consistent with this assertion, an increase in 2AP fluorescence in the second transition is attributed to the unfolding of the residual hairpin pkHP1, with concomitant loss of 2AP stacking interactions. Again, not much change in 2AP fluorescence is expected in the unfolding of alternative pkHP2 hairpin, with 2AP located in the loop of that hairpin.

A two-step van't Hoff analysis of the fluorescence melting profile of VPK-2AP revealed $T_{m1} \approx 49$ °C and $T_{m2} \approx 83$ °C (SI Table S1). Therefore, while the $T_{m1}$ values of VPK-2AP obtained from absorbance and fluorescence melting profiles are reasonably consistent, the lower $T_{m2}$ value observed in fluorescence measurements is attributed to partial unstacking of 2AP from stem 1 of the pkHP1 intermediate as a result of "pre-melting" or "end-fraying" of that stem prior to complete unfolding. Similar discrepancies between 2AP fluorescence and absorbance melting profiles have also been observed in previous studies of DNA hairpins (35).

To further examine whether PK→pkHP1→U is indeed the dominant unfolding pathway at 50 mM KCl, we measured the fluorescence melting profiles of VPK-pC. Strikingly, this melting profile exhibited only one dominant transition (Figures 2e-f), with $T_m \approx 43$ °C consistent with the unfolding of the pseudoknot. The unfolding of any residual hairpins expected in the second melting transition, as seen in absorbance measurements on this construct, are undetected by the pC probe located in stem 2. Instead, the temperature dependence of pC fluorescence before and after the transition reflects the intrinsic change in the quantum yield of the pC fluorophore, as measured in a 5-nt long reference sample, pC-ref, which is used as a control (SI Figure S3).

We expect the pC fluorescence in VPK-pC to increase upon melting of stem 2, from loss of stacking interactions (55), and to stay largely unperturbed upon melting of stem 1, either in the PK→pkHP2 step (Figure 2d; bottom pathway) or in the pkHP1→U step (Figure 2d; top pathway). The observation that pC fluorescence increases in a single transition at low but not at high temperatures, further reaffirms that nearly all of stem 2 is lost when the pseudoknot unfolds. Altogether, the fluorescence melting profiles of VPK-2AP and VPK-pC strongly support the conclusion that the dominant pathway for VPK unfolding populates pkHP1 as the intermediate state, with an alternative pathway involving pkHP2 playing only a minor role, at low ion concentrations.

**2AP-probed thermodynamics and folding/unfolding kinetics measured on hairpin HP1:** To isolate the thermodynamics and folding/unfolding kinetics of the pkHP1 ⟺ U transition, we studied a shorter sequence with the same nucleotide composition (nucleotides 1-23) as stem 1 and loop 1 of VPK (HP1-2AP in Figure 1). Nucleotides 20–23 were added at the 3' end of stem 1 to approximately mimic the dangling end in the pkHP1 hairpin conformation, and 2AP was incorporated at the same position 20 as in VPK. The fluorescence melting profiles of HP1-2AP reveal an increase in 2AP fluorescence in a single (one-step) unfolding transition with $T_m \approx 84$ °C (Figure 3b and SI Figure S4), similar to the 2AP-probed melting temperature of the second transition in VPK-2AP, $T_{m2} \approx 83$ °C (SI Table S1). These results further support our conclusion that the second transition in VPK unfolding corresponds to the melting of stem 1 when the intermediate pkHP1 unfolds.

To measure the folding/unfolding kinetics of HP1-2AP, we collected relaxation traces in response to a ~$3 - 10$ °C T-jump perturbation by recording 2AP fluorescence as a function of time (Figures 3c-f). We also measured a 5-nt reference sample, 2AP-ref, with the 2AP probe placed in the middle; this short RNA strand is not expected to exhibit any folding/unfolding kinetics and hence served as a control (SI Figure S5). T-jump measurements on 2AP-ref showed a rapid initial drop (faster than our earliest observable time of 30 μs), which reflects a drop in 2AP quantum yield as a consequence of the T-jump, followed by a slow relaxation on >10 ms time-scales that describes the decay ("recovery") of the temperature of the heated volume back to its initial pre-T-jump value (SI Figure S5b; see also SI Methods 1.7).

In contrast to control measurements, T-jump measurements on HP1-2AP showed a rapid *increase* in 2AP fluorescence immediately after the T-jump, followed by a further increase in the time-window of 30 μs $- 1$ ms, and then a decrease in 2AP fluorescence back to the pre-T-jump value on 10–300 ms time-window (Figures 3c-f). The initial increase in the 2AP-probed kinetics immediately after the T-jump is consistent with equilibrium measurements on HP1-2AP, which exhibit a gradual increase in 2AP fluorescence between 20–60 °C, prior to the melting transition, ascribed to increased unstacking of 2AP from the hairpin's stem as the temperature is raised (SI Figure S4a). In T-jump measurements, this "pre-melting" is too fast to be resolved and appears as a sudden jump in 2AP fluorescence. The relaxation kinetics observed in the 30 μs $- 1$ ms time-window is attributed to hairpin folding/unfolding kinetics, while the slower relaxation on time-scales >10 ms mimics the T-jump "recovery" decay profile seen in the 2AP-ref control measurements (SI Figure S5b).

**Global analysis of thermodynamics and folding/unfolding kinetics of HP1-2AP:** We next analyzed the equilibrium and kinetics data on HP1-2AP in terms of a two-state system: HP1 $\Longleftrightarrow$ U (Figure 3a), using a global analysis approach as described in SI Methods 1.8. The equilibrium populations of the two states at each temperature were determined in terms of two parameters: the melting temperature and the enthalpy change between the hairpin and the unfolded state; the temperature dependence of the fluorescence of each state was parameterized in terms of linear baselines. The relaxation traces were described in terms of the time-dependent change in the populations as a function of time. To this end, the rate coefficients for the hairpin to unfolded transition at each temperature ($k_{HP1 \to U}$) were determined by two parameters: the rate coefficient ($k_{HP1 \to U}^0$) at a reference temperature ($T^0$), chosen to be 85 °C, and the enthalpy barrier ($\Delta H_{HP1 \to U}^{\ddagger}$) for that transition; the folding rate coefficients at each temperature ($k_{U \to HP1}$) were obtained from the unfolding rate coefficients and the equilibrium parameters, as described in SI Methods 1.8. Additionally, the size of the T-jump and the recovery time constants were varied independently for each kinetics trace. The best fit results from this global analysis are shown in Figure 3, and the equilibrium and kinetics parameters are summarized in SI Table S2.

**Global analysis of thermodynamics and folding/unfolding kinetics of VPK-2AP:** T-jump kinetics traces on VPK-2AP were obtained in a manner similar to HP1-2AP. Kinetics traces obtained at temperatures below ~60 °C (Figures 4d-i) displayed a sharp drop in 2AP fluorescence immediately after the T-jump, resembling the behavior of the 2AP-ref control (compare Figures 4d–i with SI Figure S5b); relaxation kinetics were observed in the time-window 30 µs–10 ms, followed by T-jump recovery kinetics on time-scales >10 ms, as before. In contrast, at temperatures above ~60 °C (Figures 4j-l), the traces are analogous to the T-jump response of HP1-2AP (compare Figures 4j–l with Figures 3c–f). These results complement thermodynamics experiments and suggest that T-jump measurements in the temperature range 30–60 °C report on the folding/unfolding kinetics of the pseudoknot to the intermediate hairpin conformations, while measurements in the 60–80 °C temperature range report on the folding/unfolding kinetics of the residual hairpins.

The combined evidence from the equilibrium melting experiments on VPK-2AP, VPK-pC, and HP1-2AP showed that the VPK pseudoknot in 50 mM KCl unfolds primarily via a pathway that populates pkHP1 as an intermediate. With this insight, we described the folding/unfolding of VPK in terms of a minimal kinetic model with 3 states: the fully folded pseudoknot (PK), the intermediate hairpin (pkHP1), and the completely unfolded state (U) (Figure 4a). Using a strategy similar to that described for HP1-2AP, we analyzed the 2AP-probed equilibrium melting profile of VPK-2AP and all the kinetics traces measured over a range of temperatures in terms of a minimal set of parameters, as outlined below.

The equilibrium populations of the three states at all temperatures were determined in terms of four parameters: the melting temperature and the enthalpy change for the PK⟺pkHP1 transition and the corresponding parameters for the pkHP1⟺U transition. The temperature dependencies of the fluorescence in each state were modeled as linear baselines, as explained in SI Methods 1.8. To characterize the kinetics traces, the time-dependent populations of the three 3 states were obtained from a solution to a master equation, with the unfolding rate coefficients for each transition, $k_{PK \to pkHP1}$ and $k_{pkHP1 \to U}$, determined in terms of two parameters each: the rate coefficients $k_{PK \to pkHP1}^{0}$ (at reference temperature of 50 °C) and $k_{pkHP1 \to U}^{0}$ (at reference temperature of 85 °C), and their corresponding enthalpy barriers ($\Delta H_{PK \to pkHP1}^{\ddagger}$ and $\Delta H_{pkHP1 \to U}^{\ddagger}$); the folding rates were determined from the unfolding rates and the equilibrium parameters as before. To constrain the parameter space, the equilibrium, kinetics and fluorescence baseline parameters for the pkHP1⟺U transition in VPK-2AP were constrained to be close (though not identical) to the corresponding values obtained from the analysis of the HP1-2AP data.

The results from this global analysis are shown in Figure 4, and the best fit parameters are summarized in SI Table S2. Reassuringly, the melting parameters obtained from the global analysis are consistent with the parameters obtained from the van't Hoff analysis of the equilibrium-only data, as indicated by the reasonably good agreement between the temperature dependence of the equilibrium population of the 3 states obtained from the two descriptions (Figure 4c). The temperature-dependence of all the rate coefficients in this 3-state model, obtained from the global analysis, are represented in SI Figure S6.

**VPK-pC melting profiles reveal evidence of a second pathway at high salt concentration:** The results at ionic conditions of 50 mM KCl support pseudoknot unfolding through one major pathway having pkHP1 as an intermediate. The strongest evidence for this conclusion came from the melting profile of VPK-pC, with the pC probe located in stem 2; in 50 mM KCl, VPK-pC exhibited only a single dominant transition at low temperatures, that indicated complete loss of stem 2 in that transition (see Figures 2e-f). However, our previous computational studies uncovered parallel pathways involving both pkHP1 and pkHP2 at 1 M NaCl (18). To examine this apparent discrepancy between experiments and simulations, we investigated the effect of varying the salt concentration on the thermodynamics of pseudoknot unfolding. To this end, we monitored the fluorescence melting profiles of VPK-pC at 100 mM, 200 mM, 500 mM and 1 M NaCl. We anticipated that evidence of a parallel pathway emerging at the higher salt conditions, via the intermediate pkHP2, should be reflected in the appearance of a second melting transition at high temperatures, corresponding to melting of stem 2 when pkHP2 unfolds.

The thermodynamics data on VPK-pC measured at different NaCl concentrations are shown in Figure 5. The melting profile of VPK-pC in 100 mM NaCl (Figure 5a) resembles that in 50 mM KCl (Figure 2e), with a predominantly single transition flanked by baselines that reflect the decrease in intrinsic pC fluorescence with increasing temperature. However, for the melting profiles measured at higher salt concentrations, the slope of the high temperature baseline decreases and eventually even changes sign, becoming positive at 1 M NaCl (Figure 5g). The change in the high temperature slope, as the salt concentration is increased, is indicative of a structural transition in the RNA involving base unstacking and/or breaking of base pairs as sensed by the pC probe. These observations match the expectation that pC fluorescence will increase upon unfolding of stem 2, pointing to the presence of a parallel unfolding pathway with pkHP2 as intermediate. Further support for these conclusions comes from the corresponding derivatives plots, where hints of a transition at high temperatures are more readily discerned (Figure 5, lower panels). Interestingly, the derivatives data also show that at salt concentrations of 100 mM–500 mM NaCl (not so clear at 1 M NaCl), two peaks are observed in the transition at lower temperatures (as opposed to one dominant peak at 50 mM KCl), indicating at least two steps in the unfolding of the pseudoknot to the intermediate hairpin states. One step is readily attributed to the PK⇔pkHP1 unfolding transition, with an increase in pC fluorescence. On the other hand, the second step could well be the parallel PK⇔pkHP2 unfolding transition, with unanticipated fluorescence changes in pC associated with this step (see Figure 2d). The data with evidence of two steps in the pseudoknot melting transition is another indicator of a parallel pathway for VPK unfolding as the salt concentration is increased above 50 mM.

**Simulations reveal parallel pathways depending on the salt concentration:** In order to support the experimental findings and obtain structural insights, we conducted molecular dynamics simulations using a coarse-grained RNA model (56) which is based on the three interaction site model introduced to simulate nucleic acids (7) (see SI Methods 1.9). Prior to performing kinetics simulations, we did equilibrium simulations at 50 mM and 1 M salt concentrations, varying the temperature by the replica-exchange method (57). Calculated heat capacities (Figure S7a–b) show two distinct transitions at 47 °C and 84 °C (50 mM), and 70 °C and 88 °C (1 M), respectively. These are in excellent agreement with previous experiments (17, 58) as well as absorbance melting profiles in this study (SI Table S1). From the equilibrium ensemble, we also obtained temperature-dependent populations of the four states. The population profile at 50 mM (Figure S8a) closely resembles the one acquired through experimental analysis except that we can detect a small population of pkHP2 together with the other three states in the simulation ensemble. The consistency between our experiments and simulations has been validated by these heat capacity and population analyses. At 1 M salt (Figure S8b), the population of pkHP1 not only shifts toward higher temperature but also decreases. Alternatively, the pkHP2 state is more populated than at 50 mM.

Next, we performed a series of kinetic simulations to reveal the details of the folding pathways. We generated more than 200 folding trajectories at various concentrations of monovalent salt. Each simulation was started from an unfolded conformation prepared at 120 °C and the temperature was quenched to 37 °C so that the folding reactions took place regardless of the salt concentration. After the trajectories reached the folded PK state, we analyzed in each trajectory whether stem 1 folded first (pkHP1 pathway) or stem 2 folded first (pkHP2 pathway). As a result, at all salt concentrations examined, the pkHP1 pathway was dominant whereas some fraction of the pkHP2 pathway was also observed (Figure 6). At 50 mM salt, 90% of folding trajectories undergo the pkHP1 pathway. At higher salt concentrations ([M+] > 500 mM), the pathway through pkHP1 becomes less dominant (~80%). This is consistent with the results of VPK-pC measurements at higher salt concentrations (Figure 5). The excellent agreement between simulations and experiments shows that the probes do not greatly alter the folding patterns of the RNA.

**DISCUSSION**

We have investigated the thermodynamics and folding/unfolding kinetics of a pseudoknot using fluorescent nucleotide analogs and molecular simulations to probe salt-dependent changes in the folding pathways. Altogether, the experiments and simulations provide a consistent description of VPK folding mechanisms and reveal how the ionic conditions modulate the flux through different folding pathways.

**Two-step assembly of VPK:** At low ionic conditions (50 mM KCl), VPK fold/unfolds predominantly by a single pathway that has two distinct transitions. In the first transition, single-stranded RNA folds to an intermediate conformation, which we identify as a hairpin structure pkHP1, in which stem 1 of the pseudoknot is formed but not stem 2. This intermediate structure takes ~170–430 μs to fold from a fully unfolded PK at 37 °C (SI Table S2). These folding times are consistent with our results on the truncated HP1 sequence (Figure 1), which forms the same hairpin structure as pkHP1 but without the long dangling end. The folding times for the isolated HP1 were found to be ~40–280 μs at 37 °C (SI Table S2) under identical ionic conditions. The slightly longer times for pkHP1 in comparison with HP1 may be due to the presence of the long dangling end in pkHP1, which could present increased intrachain interactions and thus slow down the overall folding (59). Nevertheless, the similarity in folding time of pkHP1 and the truncated HP1 shows that the species sampled at low KCl concentration is the structure in which the more stable hairpin is formed.

In the second transition, pkHP1 folds to a stable PK, with folding rates faster than the unfolding rates below ~50 °C. This behavior contrasts with the folding of some riboswitches that exist predominantly in hairpin conformations and only fold to stable H-type pseudoknot upon ligand binding (40), suggesting an important role for stability in the functioning of the VPK pseudoknot. At 37 °C, the folding time for the pkHP1 to PK step is ~5–13 ms (SI Table S2); this time is 5–10-fold slower than our previously reported value of 1 ms for the unlabeled VPK pseudoknot, which was obtained from time-resolved absorbance measurements (17). Comparison of the thermodynamics of unlabeled VPK with VPK-2AP, using absorbance measurements for both, shows that incorporation of the 2AP label significantly increases the stability of the intermediate hairpin structure pkHP1; the melting temperature for the unfolded-to-hairpin transition is found to be ~10 °C higher in VPK-2AP than in the unlabeled VPK (SI Table S1). This increased stability could arise from stronger stacking interactions between stem 1 and 2AP at position 20 (in VPK-2AP) than the corresponding interactions between stem 1 and adenine (in unlabeled VPK). Therefore, a likely scenario consistent with our thermodynamics and kinetics results is that the transition from pkHP1 to the folded pseudoknot conformation requires partial unstacking of the nucleotide at position 20 that prefers to be stacked over stem 1 in the intermediate hairpin structure. This unstacking would then be energetically more costly for the more stable intermediate pkHP1 conformation in VPK-2AP. Hence, the folding of VPK-2AP would be correspondingly slower than the folding of unlabeled VPK.

**Fluxes through the dominant pathways depend on salt concentration:** The major finding in our study is that the flux between the two pathways changes as the concentration of monovalent cation is altered. There is a predominantly single folding pathway to form the pseudoknot structure in 50 mM KCl. In contrast, our previous computational studies, carried out at 1 M NaCl, showed that VPK folds through parallel pathways, with a significant fraction (~23%, Figure 6a) of molecules folding through the alternative pathway that populates the less stable pkHP2 (18). To assess the role of ionic strength in modulating the folding pathways partition, we performed equilibrium experiments on VPK-pC, which directly probes the folding/unfolding of stem 2, at various salt concentrations; these measurements indeed uncovered signatures of the appearance of an alternative unfolding pathway as the ionic strength was increased from 100 mM to 1 M NaCl (Figure 5).

Our simulations covering lower ionic strength conditions shows, in agreement with the experiments, that the fraction of molecules that folds via a pkHP2 pathway diminishes as we decrease the monovalent salt concentration, from ~20% at 1 M to ~10% at 50 mM monovalent salt (Figure 6b). At 37 °C, the simulations predict that the folding from the unfolded (U) state to pkHP1 occurs at ~70 μs, and that the folding from pkHP1 to the folded PK structure occurs at ~1.5 ms (SI Table S3). Both folding times are smaller than the folding times obtained from experiments on VPK-2AP (~270 μs and ~8 ms, respectively). It should be emphasized that, the predicted folding times are in excellent agreement with the ~1 ms folding time previously reported for the unlabeled VPK (17). We surmise that the mismatch between our measurements on VPK-2AP and the predictions from the computational studies may be in part due to the stabilizing effect of the 2AP probe on the intermediate pkHP1 conformation. The enhanced stability of 2AP-labeled pkHP1 could also favor that pathway over the pkHP2 pathway, and account for why experiments implicate predominantly a single folding pathway while simulations indicate ~10% folding trajectories via the pkHP2 pathway even at 50 mM salt (Figure 6). We also cannot rule out the possibility that experiments and modeling may potentially miss the small (~10%) fraction of molecules that fold via the alternative pathway in VPK-2AP at 50 mM KCl. For this parallel pathway, simulations indicate that the intermediate hairpin structure folds in ~50 μs followed by folding to the pseudoknot in ~110 μs (SI Table S3). Thus, the total folding time through this pathway is faster than through the predominant one, suggesting that at the given ionic condition, folding is thermodynamically and not kinetically controlled.

**Assembly order is determined by stability of hairpins:** Based on simulations of several PKs, Cho, Pincus, and Thirumalai (CPT) showed that assembly mechanism of PKs is largely determined by stabilities of the constituent secondary structural elements, which are pkHP1 and pkHP2 (Figure 2) (18). Their stabilities could be changed by sequence as well as ionic conditions. With this stability principle in mind, our experimental and computational results allow us to answer the questions: What causes the flux through the minor alternative pathway to increase with increasing salt? To gain insight into how the folding trajectories partition into the parallel pathways, we consider the nature of the intermediate hairpins states formed in each pathway. The pkHP1 state has a stem with 5 G-C base pairs and a loop of 9 nucleotides, while pkHP2 has a stem of 6 base pairs, 4 G-C, 1 A-U and 1 G-U, and a loop of 14 nucleotides. Based on the composition of the two stems and the corresponding loop lengths, pkHP1 is more stable than pkHP2. The stem of pkHP1 has a larger G-C content and is therefore more stable than the stem of pkHP2; additionally, from entropic considerations, the smaller loop of pkHP1 is more stable than the loop of pkHP2 (60). Thus, the stability principle (18) would predict – as observed in experiments and simulations – that the dominant pathway ought to be through pkHP1.

What happens to each of these structures as we increase the salt? Previous studies on RNA hairpin stability at different salt conditions showed that the stability of the RNA hairpins stems is not significantly affected as the monovalent ion concentration is increased, but the stability of the loops increase (61). Furthermore, thermodynamic measurements on ssDNA and RNA hairpins of varying loop sizes demonstrated that smaller loops are significantly more stable compared with larger loops than expected from entropic considerations, especially in the presence of 100 mM monovalent ions (60, 62). Moreover, computational studies on nucleic acid hairpin folding showed that the difference in stability between hairpins with small and large loops is further amplified as the salt concentration decreases (63). In light of these results, we therefore conclude that, while

pkHP1 is more stable than pkHP2 at all salt conditions, the difference in the stabilities between the two hairpins decreases as we increase the concentration of monovalent ions, thereby increasing the probability that some fraction of VPK molecules will fold via the alternative pkHP2 pathway (Figure 7). Thus, altering the ionic strength or changing the sequence of the hairpins, which changes the relative stabilities of the hairpin stems and/or loops, will result in change in fluxes through the pathways. This demonstration in the present combined experimental and computational study is in complete accord with the stability principle for PK assembly enunciated by CPT.

It is interesting to contrast the unambiguous demonstration that VPK folds by parallel pathways to the difficulties in showing similar behavior in the folding of small single domain proteins. The flux through a pathway, say II, relative to a competing route I, is $\Phi_{II} \propto e^{-(G_{II}-G_I)/k_B T}$. The magnitude of the quantity $(G_{II} - G_I)$, related to the stability difference between pkHP1 and pkHP2 in the intact PK, has to be such that $\Phi_{II}$ can be measured or inferred, as we have done here for the VPK pseudoknot (Figure 6b). In several RNA PKs the difference between $G_{II}$ and $G_I$ can be altered by changing sequence or ionic strength, as we demonstrate here, while in the folding of small single domain proteins it has not been straightforward to tune the folding landscape parameters in order to modulate the flux through alternate pathways.

**CONCLUSIONS**

Our experiments and modeling provide a remarkably self-consistent picture of RNA pseudoknot folding mechanisms and also provide insight into the role of monovalent ion concentrations in partitioning into different folding pathways. It would be most instructive to consider PK folding in the presence of divalent ions, which could have a significant effect on the stabilities of the intermediate states and on the folding pathways, through specific ion-RNA interactions. Further studies that synergistically combine experiments and simulations, as in this study, are needed to elucidate the nuances of folding pathways and mechanisms for these and other RNA pseudoknots, which in turn will be invaluable for understanding the folding of larger and more complicated ribozymes.

## METHODS

**RNA samples.** Unlabeled RNA samples were obtained from Dharmacon Inc., CO. 2-aminopurine (2AP) and pyrrolocytidine (pC) labeled samples were obtained from TriLink Biotechnologies, CA. Samples were prepared as described in SI Methods 1.1. All measurements were made in 10 mM MOPS buffer pH 7.0, with either 50 mM KCl or a range of NaCl concentrations from 100 mM to 1 M NaCl.

**Equilibrium melting experiments.** Absorbance and fluorescence melting profiles were measured and analyzed as described in SI Methods 1.2–1.4.

**Temperature jump experiments**. Kinetic measurements were carried out using a home built laser temperature jump (T-jump) apparatus, which uses 10-ns laser pulses at 1550 nm, generated by Raman shifting the 1064 nm pulses from the output of an Nd:YAG laser, to rapidly heat a small volume of the sample within ~10 ns (60, 64). Details of the T-jump apparatus and analysis of the relaxation traces are in SI Methods 1.5–1.7.

**Global analysis of equilibrium and T-jump relaxation traces.** The equilibrium and kinetics data were simultaneously analyzed in terms of a minimal kinetic scheme, using a master-equation approach, as described in SI Methods 1.8.

**Simulation model and protocols**. We performed molecular dynamics simulations using a coarse-grained RNA model. The model and parameters were calibrated and described in details elsewhere (56). Further details are in SI Methods 1.9.

## ACKNOWLEDGEMENTS

We thank Jie Liang and Ke Tang for many helpful discussions. We also thank Alex Sarmiento for his extensive help in the implementation of the global modeling algorithms. We acknowledge the Texas Advanced Computing Center at The University of Texas at Austin for providing computational resources. This work was supported in part by grants from the National Science Foundation (MCB−1158217 to A.A. and CHE 16−36424 to D.T.). D.T. also acknowledges additional support by the Collie-Welch Regents Chair (F-0019).

**FIGURE LEGENDS**

**Figure 1.** RNA constructs used in this study. Pseudoknot VPK samples were labeled either with 2-aminopurine at position 20 (VPK-2AP) or with pyrrolocytidine at position 31 (VPK-pC). Hairpin HP1-2AP is a truncated sequence consisting of nucleotides 1−23 of VPK-2AP. Colors represent different regions of the pseudoknot: blue, stem 1; yellow, loop 1; green, stem 2; magenta, junction; and red, loop 2. In HP1-2AP, nucleotides 8−14 of VPK stem 2 become part of the loop in this truncated hairpin structure.

**Figure 2.** Thermodynamics of VPK melting from fluorescence experiments. (a,d) Folding/unfolding schemes for (a) VPK-2AP and (d) VPK-pC. (b,e) 2AP fluorescence of (b) VPK-2AP and (e) pC fluorescence of VPK-pC are plotted as a function of temperature. (c,f) The corresponding first derivatives of the fluorescence with respect to temperature ($\delta F / \delta T$) are plotted as a function of temperature for (c) VPK-2AP and (f) VPK-pC. The data (symbols) are the averages of 2−3 independent sets of measurements. The error bars represent the standard deviation of the mean. For clarity, only every other data point is presented. The continuous lines are fits to two sequential van't Hoff transitions.

**Figure 3.** Global analysis of thermodynamics and kinetics measurements on HP1-2AP. (a) A minimal two-state folding/unfolding scheme for HP1-2AP: the fully folded hairpin state (HP1) and the fully unfolded state (U). (b) Fluorescence melting profile of HP1-2AP. The data (symbols) are the averages of two independent sets of measurements. The error bars represent the standard deviation of the mean and are smaller than symbols. For clarity, only every other data point is presented. The continuous black line is from a global fit of the equilibrium and kinetics data to a 2-states kinetic model, as described in the text; the gray line represents the temperature dependence of 2AP fluorescence in the hairpin state, as obtained from this global fit. (c−f) Relaxation traces from T-jump measurements on HP1-2AP are shown for four different sets of initial and final temperatures, with an average T-jump $\Delta T = 4.2 \pm 0.8$ °C. Each trace shows the fluorescence intensity level at the initial temperature, prior to the arrival of the IR pulse (denoted by negative times on the $x$-axes), a sharp increase in fluorescence immediately after the IR pulse that heats the sample, reflecting an increase in 2AP-fluorescence, possibly from rapid unstacking, followed by relaxation kinetics in the time-window of ~30 μs to ~1 ms, and then the decay of the temperature

of the sample back to the initial temperature, on time-scales >10 ms. The continuous black lines in each of the panels are from the global fit of the equilibrium and kinetics data.

**Figure 4.** Global analysis of thermodynamics and kinetics measurements on VPK-2AP. (a) A minimal three-state folding/unfolding scheme for VPK-2AP: the fully folded pseudoknot (PK), a partially folded hairpin state with stem 1 formed but not stem 2 (pkHP1) and unfolded RNA (U). VPK-2AP fluorescence equilibrium and kinetics data were simultaneously fitted using this model. (b) The fluorescence melting profile of VPK-2AP; the symbols are as described in Figure 4a. The continuous black line is from a global fit of the equilibrium and kinetics data to a 3-states kinetic model, as described in the text; the pink (green) line represents the temperature dependence of 2AP fluorescence in the PK (pkHP1) state, as obtained from this global fit. (c) Equilibrium populations of the three states indicated in panel (a) are plotted as a function of temperature: PK (pink); pkHP1 (green); U (blue). The populations are calculated from either the parameters from the global kinetic modeling (continuous lines) or from the van't Hoff analysis (dashed lines, SI Table S1). (d–l) Relaxation traces from T-jump measurements on VPK-2AP are shown for nine different sets of initial and final temperatures, with an average T-jump $\Delta T = 9.1 \pm 3.5$ °C. The continuous black lines are from a global fit of the equilibrium and kinetics data.

**Figure 5.** Salt dependence of VPK-pC melting profiles. (Top) Fluorescence of VPK-pC at four different NaCl concentrations; (bottom) the corresponding first derivatives $\delta F / \delta T$ are plotted as a function of temperature. The data (symbols) are the averages of two independent sets of measurements. The error bars represent the standard deviation of the mean. For clarity, only every other data point is presented.

**Figure 6.** Parallel folding pathways of the VPK pseudoknot. (a) The states of VPK are designated as PK: folded pseudoknot; pkHP1 and pkHP2: partially folded (hairpin) states; U: unfolded state. In the top pathway, stem 1 folds first to form the pkHP1 hairpin state; in the bottom pathway, stem 2 folds first to form the pkHP2 hairpin state. The structures representing the four states were taken from simulations. (b) Salt-dependence of the VPK folding pathways from coarse-grained simulations. The fraction of molecules that fold via the pathway that populates pkHP1 as an intermediate, is plotted as a function of the monovalent salt concentrations. The error bars represent the 95% exact confidential intervals. The free energy difference between the two pathways, calculated based on the fraction of pkHP1, is marked on the right axis.

**Figure 7.** Modulation of the folding pathways of the VPK pseudoknot by monovalent ions. At low salt concentration, the pseudoknot has a dominant folding pathway, with the hairpin pkHP1 as an intermediate state. As the salt concentration increases, the hairpin pkHP2 becomes stable causing the emergence of a parallel folding pathway.

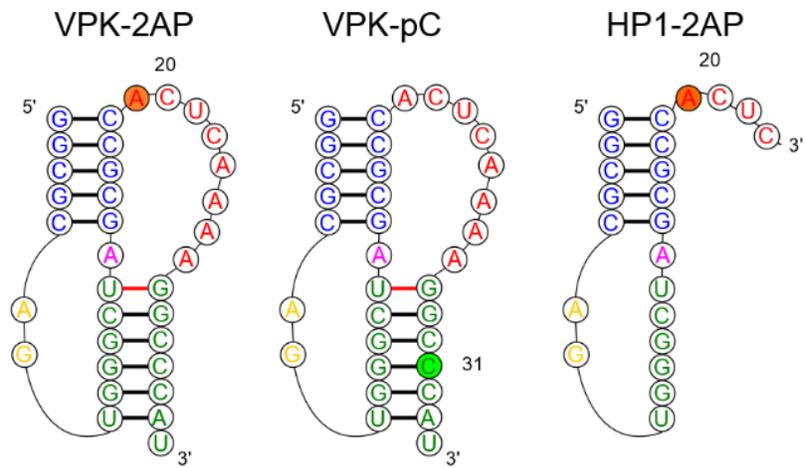

**Roca et al.**
**Figure 1**

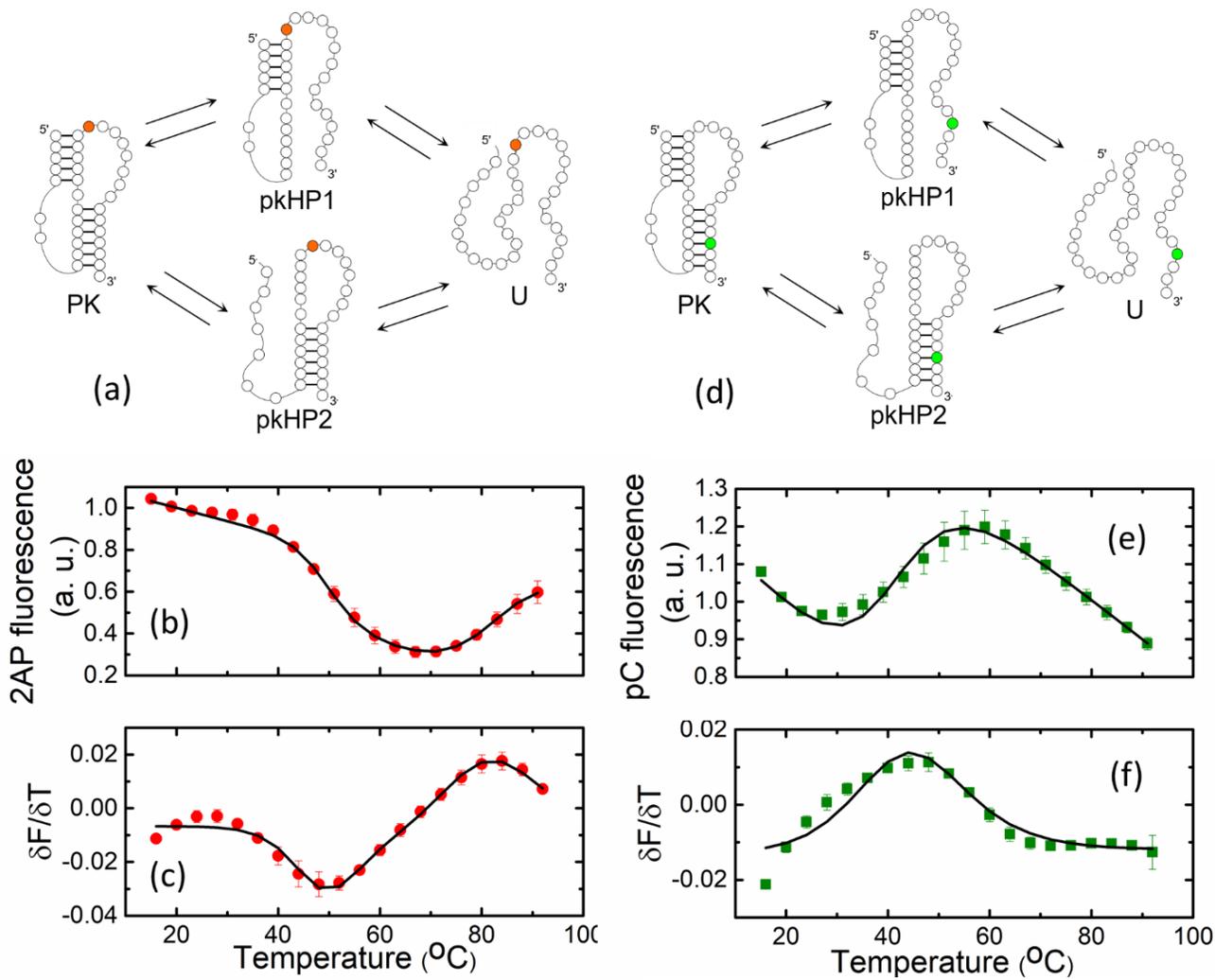

(a)

pkHP1

U

PK

pkHP2

(b)

(c)

(d)

pkHP1

U

PK

pkHP2

(e)

(f)

Roca et al.
Figure 2

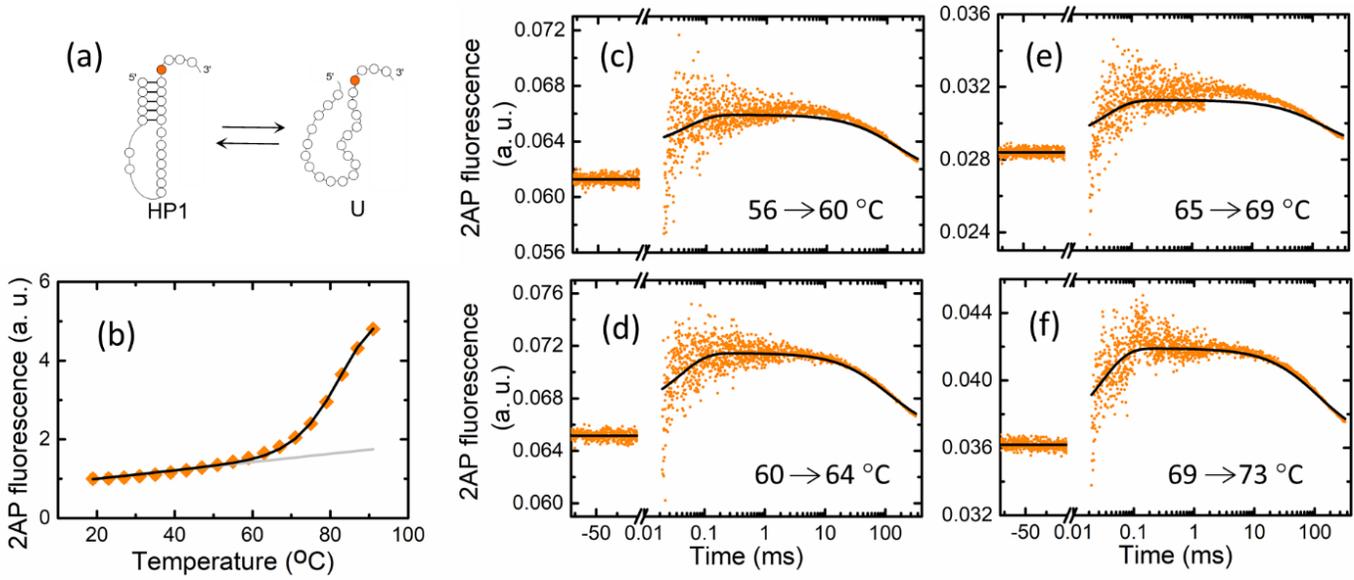

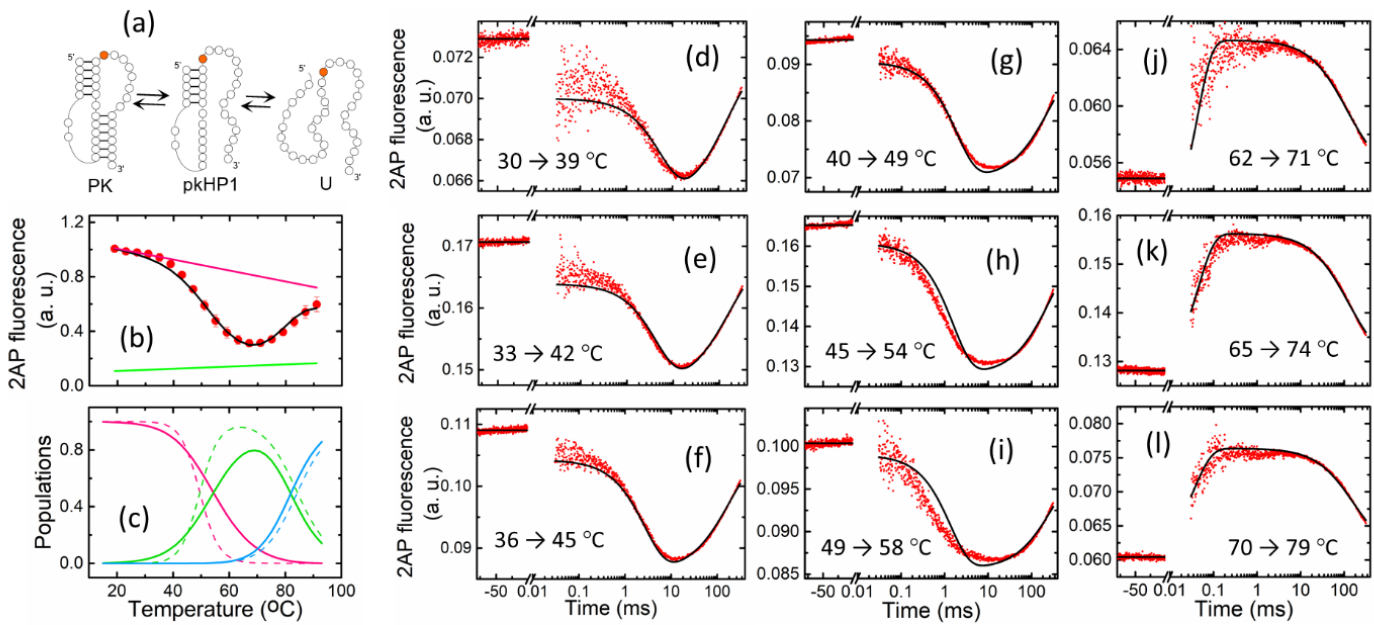

Roca et al.

Figure 4

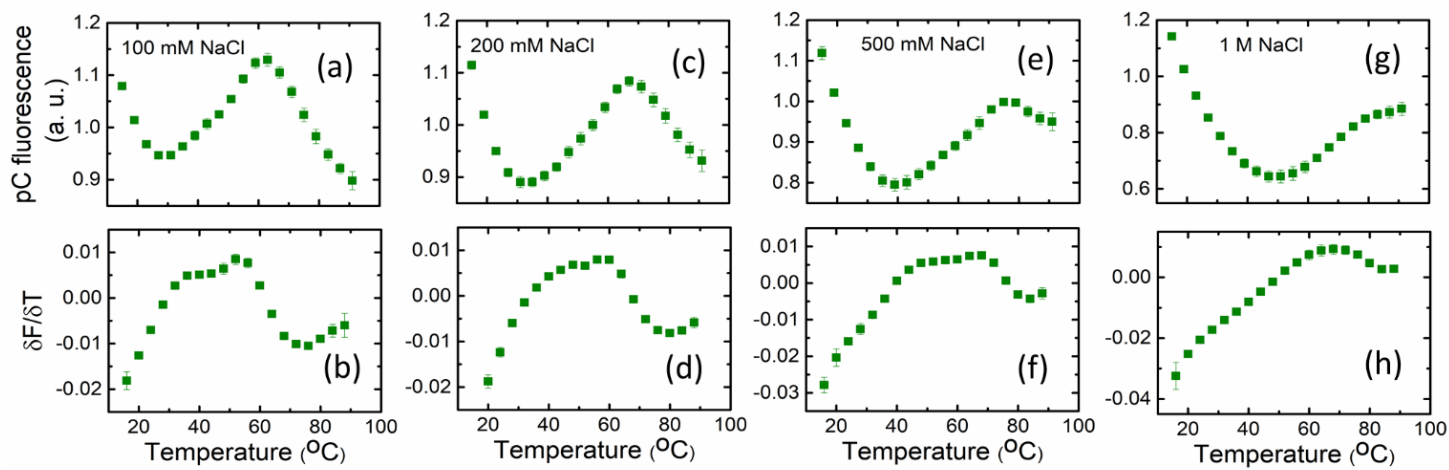

Roca et al.
Figure 5

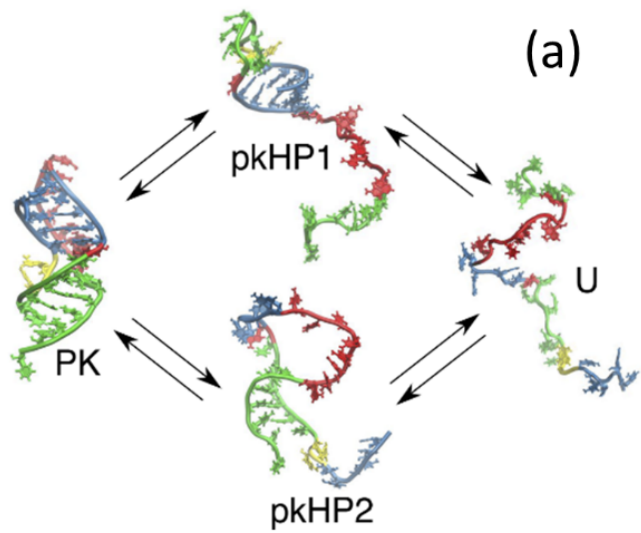

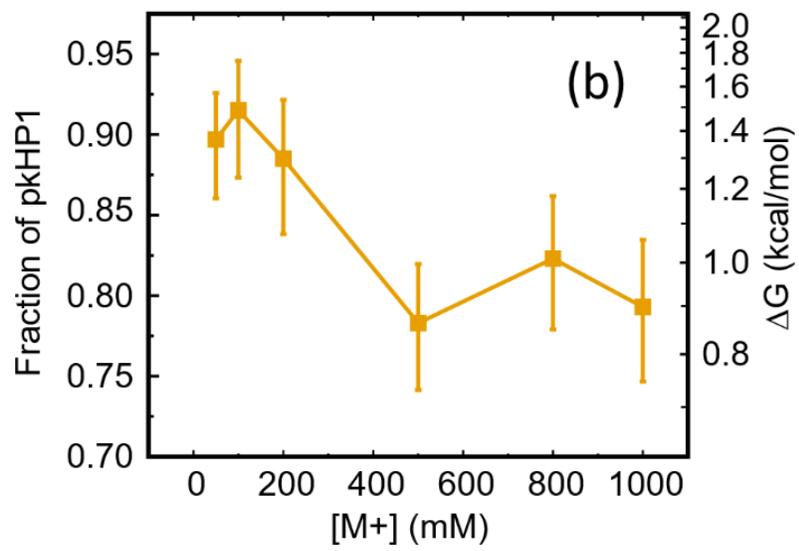

Roca et al.
Figure 6

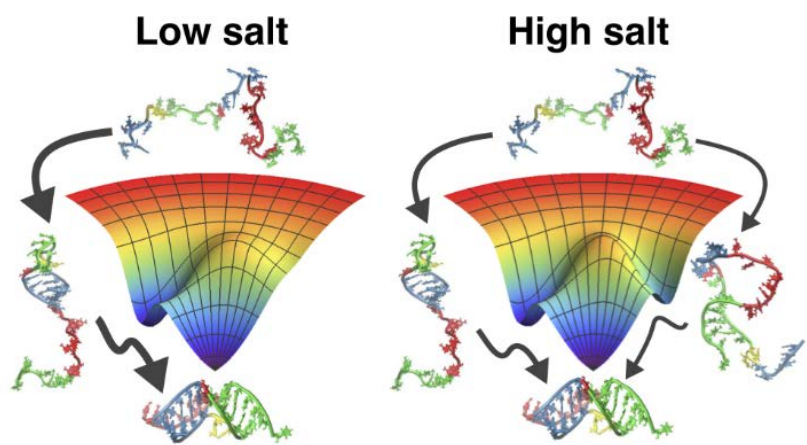

**Roca et al.**
**Figure 7**

**Supporting Information**

Monovalent ions modulate the flux through multiple folding pathways of an RNA pseudoknot

Jorjethe Roca[1#], Naoto Hori[2], Yogambigai Velmurugu[1], Ranjani Narayanan[1‡], Prasanth Narayanan[1], D. Thirumalai[2], and Anjum Ansari[1,3*]

[1]Department of Physics and [3]Department of Bioengineering,

University of Illinois at Chicago, Chicago, IL 60607

[2]Department of Chemistry, University of Texas at Austin, Austin, TX 78712

[#]Present address: Thomas C. Jenkins Department of Biophysics, Johns Hopkins University,

Baltimore MD 21218

[‡]Present address: Division of Physics & Applied Physics, School of Physical and Mathematical

Sciences, Nanyang Technological University, Singapore 637371

[*]Corresponding author.

E-mail: ansari@uic.edu. Phone: (312) 996−8735. FAX: (312) 996−9016

**SI METHODS**

**1.1 RNA samples.** Unlabeled RNA samples were obtained from Dharmacon Inc., CO. 2-aminopurine (2AP) and pyrrolocytidine (pC) labeled samples were obtained from TriLink Biotechnologies, CA. In this study, two RNA constructs were used: the pseudoknot VPK and its constituent hairpin HP1 (Figure 1). The pseudoknot samples used were either unlabeled, labeled with 2AP at position 20 (VPK-2AP), or labeled with pC at position 31 (VPK-pC). The constituent hairpin HP1 was labeled at position 20 (HP1-2AP). All pseudoknot and hairpin samples were purchased with PAGE purification. Two short oligonucleotides, 2AP-ref (5'-CC[2AP]CU-3') and pC-ref (5'-GC[pC]CA-3'), were used as reference samples to measure the temperature dependence of the quantum yield of the 2AP and pC probes, respectively, in a sequence context identical to that in the VPK or the constituent hairpin. The reference samples were purchased with HPLC purification. All measurements were made in 10 mM MOPS buffer pH 7.0, with either 50 mM KCl or a range of NaCl concentrations from 100 mM to 1 M. Prior to measurements, the samples were unfolded by first heating to 60 °C and waiting at that temperature for 10 min, then further heating to 90 °C and waiting there for 10 more min followed by ice cooling and equilibration for another 10 min.

**1.2 Equilibrium melting experiments.** Absorbance melting profiles were obtained for VPK, VPK-2AP and VPK-pC. Fluorescence melting profiles were obtained for VPK-2AP, VPK-pC and HP1-2AP. Strand concentrations for all measurements were calculated by making absorbance measurements at 260 nm, with extinction coefficients of 323,500 $M^{-1}cm^{-1}$ for VPK, 325,100 $M^{-1}cm^{-1}$ for VPK-2AP and VPK-pC, and 212,700 $M^{-1}cm^{-1}$ for HP1-2AP, as reported by the manufacturer.

Absorbance melting profiles were collected using a UV-Vis spectrophotometer Cary50 (Varian Instruments, CA). The sample absorbance was measured in a quartz cuvette of 1 or 0.1 cm path length, with excitation at 260 nm. The temperature of the sample cuvette was controlled by a Peltier device incorporated in the instrument. Melting profiles were obtained by making absorbance measurements as a function of temperature, with the temperature increased at a rate of 0.4 °C/min from 25 °C to 93 °C. Three absorbance experiments were done for VPK: one at 10 μM strand concentration and two at 18.5 μM. Two absorbance experiments were obtained for VPK-pC at 1.7 μM and two for VPK-2AP at 12.6 μM concentration.

Fluorescence equilibrium melting profiles were obtained using a Fluoromax-4 (HORIBA Instruments Inc., NJ) spectrofluorimeter. 2AP- (pC-) labeled samples were excited at 310 (350) nm and the fluorescence emission spectra collected over a wavelength range of 320 nm – 600 nm (360 nm – 600 nm). Fluorescence emission spectra were recorded every 2 °C from 15 °C to 93 °C after an equilibration time of 5 min at each temperature. The fluorescence melting profiles were measured at strand concentrations of 5.8 μM and 63.2 μM for VPK-2AP (2 measurements at each concentration), 6.8 μM and 17.5 μM for VPK-pC, 4.2 μM and 41.8 μM for HP1-2AP, 12.5 μM and 25 μM for 2AP-ref, and 9.2 μM (2 experiments) for pC-ref. The melting profiles were constructed by plotting the maximum of the fluorescence emission spectra (at 370 nm for 2AP-labeled samples and at 460 nm for pC-labeled samples) for each measurement versus temperature. For both absorbance and fluorescence experiments, reversibility was checked by cooling the sample back to its initial temperature and immediately repeating the melting measurements.

All melting profiles were smoothed using the moving average option of the smooth built-in function of MATLAB (R2014a 8.3.0.532). Final absorbance and fluorescence melting profiles were constructed by averaging all the smoothed melting profiles obtained for a given sample (2-3 per experiment); the error was calculated as the standard deviation of the mean. These averaged melting profiles were subsequently normalized to 1 at 20 °C. The thermodynamic parameters describing the melting profiles were obtained using two methods, as described below.

**1.3 Analysis of melting profiles.** All melting profiles that exhibited a single-transition were described in terms of a van't Hoff transition with linear upper and lower baselines, and two parameters describing the transition: $\Delta H_{vH}$, the enthalpy difference between the folded and the unfolded conformations, and $T_m$, the melting temperature defined by the mid-point of the transition. Melting profiles that exhibited two transitions were described in terms of two sequential van't Hoff transitions, with three baselines and four parameters: $\Delta H_{vH1}$, $\Delta H_{vH2}$, $T_{m1}$ and $T_{m2}$. For the absorbance melting profiles, $A(T)$, all baselines were constrained to have the same slope. For the fluorescence melting profiles, $F(T)$, the baselines were allowed to have different slopes but the baselines corresponding to the completely unfolded state were constrained to have the same slope as the fluorescence profiles for the reference samples: 2AP-ref and pC-ref.

*Method 1*. For the more general case of two van't Hoff transitions and different baselines, the melting profiles were described as:

$$F(T) = F_1(T) + \big(F_2(T) - F_1(T)\big)f_1 + \big(F_3(T) - F_2(T)\big)f_2 \qquad (S1)$$

where $F(T)$ is the fluorescence of the sample, $F_1(T)$, $F_2(T)$ and $F_3(T)$ are the baselines associated to the three states of the melting process, parameterized as linear functions of temperature, each with different slopes (the slope of $F_3(T)$ is the same as that of the reference sample 2AP-ref or pC-ref). $f_1$ and $f_2$ are the unfolded fractions as a function of temperature, described in terms of the van't Hoff enthalpy change $\Delta H_{vH}$ and melting temperature $T_m$ as follows:

$$f_i = \left(1 + \exp\left[-\frac{\Delta H_{vHi}}{R}\left(\frac{1}{T} - \frac{1}{T_{mi}}\right)\right]\right)^{-1} \qquad (S2)$$

Therefore, the total number of parameters are 9 when choosing baselines with independent slopes (5 parameters for the three baselines and 4 parameters for the unfolding transitions). For absorbance melting profiles exhibiting two transitions, all baselines have the same slope, and thus the number of parameters is reduced to 8. Similarly, melting profiles with only one unfolding transition were fully described by 5 parameters when the slopes of the baselines were different and by 4 parameters when the slopes were the same.

*Method 2.* Alternatively, the derivative of the melting profiles $\partial F / \partial T$ can be obtained numerically from the measured $F(T)$, and described in terms of the van't Hoff three-state system as:

$$\frac{\partial F}{\partial T} = \frac{\partial F_1}{\partial T} + \frac{\partial (F_2 - F_1)}{\partial T} f_1 + \frac{\partial (F_3 - F_2)}{\partial T} f_2 + (F_2 - F_1) \frac{\partial f_1}{\partial T} + (F_3 - F_2) \frac{\partial f_2}{\partial T} \qquad (S3)$$

or:

$$\frac{\partial F}{\partial T} = m_1 + (m_2 - m_1) f_1 + (m_3 - m_2) f_2 + (F_2 - F_1) \frac{\partial f_1}{\partial T} + (F_3 - F_2) \frac{\partial f_2}{\partial T} \qquad (S4)$$

where the $m_i$'s are the slopes of the fluorescence baselines and:

$$\frac{\partial f_i}{\partial T} = -\left(\frac{\Delta H_{vHi}}{RT^2}\right) \frac{\exp\left(-\frac{\Delta H_{vHi}}{R}\left(\frac{1}{T} - \frac{1}{T_{mi}}\right)\right)}{\left[1 + \exp\left(-\frac{\Delta H_{vHi}}{R}\left(\frac{1}{T} - \frac{1}{T_{mi}}\right)\right)\right]^2} \qquad (S5)$$

In both methods, the melting parameters were obtained by fitting either the fluorescence (or absorbance) or its derivative as a function of temperature using a least-square non-linear procedure in MATLAB (R2014A 8.3.0.532) that minimizes the chi-square (1):

$$\chi^2 = \sum_{i=1}^{N} \frac{1}{\sigma_i^2} \left\{ \left( F_i - F(T_i) \right)^2 \right\} \qquad (S6)$$

where $\sigma_i$ is the uncertainty associated with the *i*-th data point, calculated as the standard error of the mean from several different measurements of the melting profiles.

The thermodynamic fitting parameters ($\Delta H_{vH}$ and $T_m$ for one-step transition and $\Delta H_{vH1}$, $\Delta H_{vH2}$, $T_{m1}$ and $T_{m2}$ for two-step transitions) for all the samples are summarized in SI Table S1. The primary source of error in the fitting parameters is from incomplete information about the temperature dependence of the linear baselines. The uncertainties in the parameters, reported in SI Table S1, were estimated as described in the next section.

**1.4 Error estimation of melting parameters.** To obtain the uncertainties of the fitting parameters in the melting analysis, we varied each fitting parameter by a fixed amount, one at the time, and refitted the melting data to minimize the chi-square by allowing the rest of the parameters to vary. From the $\chi^2$ vs parameter value plots thus generated (SI Figure S2), the uncertainties in the parameters were identified as the values at which $\chi^2$ changes by 1; this increase: $\Delta\chi^2 = 1$ corresponds to one standard deviation in parameter space (1). The lower and upper errors in the parameters are represented as sub and superscripts in SI Table S1.

**1.5 Temperature jump experiments**. Kinetic measurements were carried out using a home-built laser temperature jump (T-jump) apparatus, which uses 10-ns laser pulses at 1550 nm, generated by Raman shifting the 1064 nm pulses from the output of an Nd:YAG laser, to rapidly heat a small volume of the sample within ~10 ns (2, 3). The laser pulses are focused to ~1 mm spot size onto a 2-mm wide sample cuvette of path length 0.5 mm. Each laser pulse (~40 mJ/pulse at the sample position) yields ~3–10 °C T-jump at the center of the heated volume. Kinetics traces were measured only for the 2AP-labeled samples (VPK-2AP and HP1-2AP). The probe source for excitation of 2AP was a 200-W Hg–Xe lamp, with the excitation wavelengths selected by a broadband filter centered at 320 nm (BrightLine 320/40, Semrock, Rochester, New York). The fluorescence emission intensity was monitored using a Hamamatsu R928 photomultiplier tube equipped with another broadband filter (BrightLine 370/36, Semrock, Rochester, New York) and coupled with a preamplifier Hamamatsu C1053-51 and recorded in a 500 MHz transient digitizer (Tektronix, DPO4054B, Beaverton, OR).

The initial temperature of the sample was measured using a thermistor (YSI 44008; YSI Inc., Yellow Springs, OH) in direct contact with the sample cell. The magnitude of the temperature change upon T-jump perturbation was determined from measurements on control (2AP-ref) samples, for which no relaxation kinetics other than T-jump recovery are expected, by comparing the initial relative change in fluorescence intensity immediately following T-jump with the relative fluorescence changes expected from equilibrium temperature scan measurements of the same sample. The errors in the T-jump estimates are about 10–20%.

**1.6 Acquisition and analyses of T-jump relaxation traces.** To acquire data with the highest temporal resolution and be able to span several orders of magnitudes in time scale, it is necessary to measure T-jump kinetics traces over different time scales and then combine these traces. We typically acquired the kinetics traces on at least two time scales, with one million data points in each trace; the short time scale covered kinetics up to 2 ms, with a time-resolution of 2 ns, while the longer time scale covered kinetics up to 400 ms, with a time-resolution of 400 ns. For each time scale, 512 kinetics traces were acquired and averaged by the digitizer, and saved for further analysis. The data acquired in the short time scale were reduced to 400 points and the data in the long time scale were reduced to 2,000; in both cases a logarithmic average approach was used. In this approach, the temporal data was converted to a logarithmic scale and a given amount of points were averaged together to yield only 100 points per logarithmic decade. Given that there are more data points in time decades corresponding to longer times, the number of points averaged increased as the time increased. By averaging this way, the resolution in the smallest time scale is preserved while the high density of points in the highest time scale is reduced. Prior to any further analysis, data acquired below ~30 µs in each trace were discarded because of artifacts either from scattered infrared laser light into the photomultiplier tube, or due to cavitation effects from microbubbles in the samples (4).

In order to combine the data acquired over the two different time scales, the two traces were fitted simultaneously with a double-exponential decay convoluted with T-jump recovery kinetics, with an additional multiplicative factor applied to one of the traces to account for any systematic difference in the measured intensities for the two traces. The double-exponential decay with T-jump recovery is described by:

$$I(t) = \{[I(0^+) - I_1]\exp(-k_1 t) + [I_1 - I(\infty)]\exp(-k_2 t) - I(0^-)\}f_{rec}(t) + I(0^-) \quad (S7)$$

Here, $k_1$ and $k_2$ are the two relaxation rates, $I(0^-)$ is the intensity at the initial temperature, $I(0^+)$ is the intensity immediately after the T-jump, $I_1$ is the intensity at the end of the fast relaxation process, $I(\infty)$ is the intensity at the end of both relaxation processes, and $f_{rec}(t) = (1 + t/\tau_{rec})^{-1}$ describes the time-dependence of the decay of the T-jump itself, with $\tau_{rec}$ as a characteristic time for the temperature of the heated volume of the sample to decay back to the initial equilibrium temperature (5). The parameters that were varied when matching the two traces were $k_1$, $k_2$, $I(0^+)$, $I_1$, $I(\infty)$ and the multiplicative scale factor needed to match the two traces. Once appropriately scaled, the two data sets were combined into a single kinetic trace that covered the time range from ~30 µs to 400 ms. The combined traces were then used for all subsequent analyses. The sample concentrations for the T-jump measurements were 63.2 µM for VPK-2AP, 41.8 µM for HP1-2AP, and 54.7 µM for 2AP-ref.

**1.7 Measurements of T-jump size and recovery kinetics.** The T-jump recovery kinetics traces were obtained from measurements on control (2AP-ref samples) that were shown to exhibit only the T-jump recovery kinetics. These measurements were carried out in a time-window of up to about 400 ms. The recovery kinetics were fitted to the following recovery function:

$$I(t) = [I(0^+) - I(0^-)]f_{rec}(t) + I(0^-) \qquad (S8)$$

with $f_{rec}(t), I(0^+)$, and $I(0^-)$ as described above. The recovery time constant $\tau_{rec}$ was determined for each sample at room temperature, and was found to be $108.5 \pm 10.2$ ms for VPK-2AP and $142.7 \pm 8.4$ ms for HP1-2AP.

The magnitude of the temperature jump was estimated by comparing the ratio of the fluorescence intensity levels $I(0^+)/I(0^-)$ of the reference sample (2AP-ref), measured in the T-jump spectrometer before and immediately after the T-jump, with the ratio of the fluorescence intensity levels $F(T_{fin})/F(T_{in})$, measured for 2AP-ref under equilibrium conditions at the initial $(T_{in})$ and final $(T_{fin})$ temperatures, such that $\frac{I(0^+)}{I(0^-)} = \frac{F(T_{fin})}{F(T_{in})}$. The value of $T_{fin}$ that satisfied this condition was assigned as the final temperature in the T-jump measurements, corresponding to the temperature of the sample immediately after the IR pulse. The values of the size of the T-jump and the recovery time constant thus obtained were used as initial guesses in a global fit analysis, as described in the next section.

**1.8 Global analysis of equilibrium and T-jump relaxation traces.** To simultaneously describe the equilibrium and kinetics measurements in terms of a minimal kinetic scheme, we follow a global analysis approach described below.

The equilibrium populations of the different macrostates in the kinetic scheme as a function of temperature are written in terms of the free energy of each of the macrostates as follows:

$$P_{i_{eq}}(T) = \frac{\exp\left(-\frac{\Delta G_i}{RT}\right)}{\sum_i \exp\left(-\frac{\Delta G_i}{RT}\right)} \tag{S9}$$

where $\Delta G_i$ is the free energy difference between the $i$-th macrostate and a reference state (typically chosen as the most folded state). $\Delta G_i$ is parameterized in terms of two parameters, the enthalpy change $\Delta H_i$ and a "melting" temperature $T_{mi}$, defined as the temperature at which the $i$-th macrostate and the reference state are equally populated (i.e. the temperature at which $\Delta G_i = 0$). Thus,

$$\Delta G_i(\mathrm{T}) = \Delta H_i \left(1 - \frac{T}{T_{mi}}\right) \tag{S10}$$

The equilibrium fluorescence intensity measured as a function of temperature can then be written as:

$$F_{eq}(T) = \sum_i F_i(T) P_{i_{eq}}(T) \tag{S11}$$

where $F_i(T)$ is the fluorescence level of the molecule in the $i$-th macrostate at temperature $T$. We further parameterize $F_i(T)$ as a linear function in temperature:

$$F_i(T) = F_{io}\left[m_i\left(T - T_0^{eq}\right) + 1\right] \tag{S12}$$

The parameter $F_{io}$ in Eq. S12 describes the fluorescence level of the $i$-th macrostate at a reference temperature $T_0^{eq}$ and $m_i$ is the corresponding slope describing the temperature dependence of $F_i(T)$.

To describe the temporal change in the fluorescence signals in response to a T-jump perturbation, we use a master equation approach to express the time-dependent change in the population of each macrostate in the kinetic scheme (6). The transitions between the various macrostates in the kinetic scheme are described in terms of a set of coupled differential equations:

$$\frac{dP_i}{dt} = \sum_{j \neq i} k_{j \to i} P_j - k_{i \to j} P_i \qquad (S13)$$

where $P_i$ ($P_j$) is the population of the $i$th ($j$th) macrostate and $k_{j \to i}$ and $k_{i \to j}$ are the rates for transitions from state $j$ to state $i$ and from state $i$ to state $j$, respectively.

The matrix form of the master equation is:

$$\frac{d\mathbf{P}(t,T)}{dt} = \mathbf{M}(T) \cdot \mathbf{P}(t,T) \qquad (S14)$$

where $\mathbf{P}$ is a column vector col$(P_1, \dots, P_\Omega)$, and $\mathbf{M}$ is a $\Omega \times \Omega$ rate matrix with $M_{ij} = k_{j \to i}, i \neq j$ as the off-diagonal elements and $M_{ii} = -\sum_{j \neq 1} k_{i \to j}$ as the diagonal elements. The time-dependent solution of the rate equations yields the change in population as a function of time $\mathbf{P}(t,T)$, and is obtained by diagonalizing the matrix $\mathbf{M}$ to obtain its eigenvalues ($\lambda_i$) and eigenvectors ($\mathrm{U}_i$).

The solution to Eq. S14 can be written as:

$$\mathbf{P}(t,T) = \exp[\mathbf{M}(T)t]\,\mathbf{P}(0) = \mathbf{U}\exp[\boldsymbol{\lambda}(T)t]\,\mathbf{U}^{-1}\mathbf{P}(0) \qquad (S15)$$

where $\exp[\boldsymbol{\lambda}(T)t]$ is a $\Omega \times \Omega$ diagonal matrix with $\exp[\lambda_i(T)t]$ as the diagonal matrix elements, $\mathbf{U}$ is a $\Omega \times \Omega$ matrix consisting of the eigenvectors, and $\mathbf{P}(0)$ is the column vector representing the populations of all microstates at $t = 0$ (6). The eigenvalues and eigenvectors of the real non-symmetrical square matrices $\mathbf{M}$ were calculated using the built-in functions in MATLAB (R2014a 8.3.0.532) and the GNU Scientific Libraries (GSL 1.16) in C.

To model relaxation kinetics in response to a T-jump from an initial temperature $T_{in}$ to a final temperature $T_{fin}$, the initial populations of the macrostates immediately after the T-jump are assumed to be identical to the equilibrium populations at $T_{in}$, i.e. $\mathbf{P}(t = 0) = \mathbf{P_{eq}}(T_{in})$, and are calculated using Eq. S9. The final populations, after the relaxation is complete, are expected to be consistent with the equilibrium populations at the final temperature, i.e. $\mathbf{P}(t = \infty, T_{fin}) = \mathbf{P_{eq}}(T_{fin})$.

The temperature dependence of the rate constants describing the transitions between the different macrostates are parameterized in terms of an Arrhenius equation:

$$k_{i \to j} = k_{i \to j}^0(T^0) \exp\left[-\frac{\Delta H_{i \to j}^{\ddagger}}{R}\left(\frac{1}{T} - \frac{1}{T^0}\right)\right] \qquad (S16)$$

where $k_{i \to j}^0$ is the rate constant at a reference temperature $T^0$ and $\Delta H_{i \to j}^{\ddagger}$ is the enthalpy barrier for the transition from the $i$-th to the $j$-th macrostate. The backward rates $k_{j \to i}(T)$ are calculated from the forward rates as:

$$k_{j \to i}(T) = k_{i \to j}(T) \exp\left[-\frac{\Delta G_j - \Delta G_i}{RT}\right] \qquad (S17)$$

To model the relaxation kinetics, we first identify the intensity levels measured in the T-jump spectrometer prior to the arrival of the infrared heating pulse, $I(0^-)$, as characteristic of the fluorescence levels of the equilibrium population at the initial temperature. To match the fluorescence levels measured in the T-jump spectrometer with those measured in the equilibrium fluorescence measurements, we use a scale factor $\varsigma$ as follows:

$$I(0^-) = \varsigma F_{eq}(T_{in}) = \varsigma \sum_i F_i(T_{in}) P_{i_{eq}}(T_{in}) \qquad (S18)$$

The scale factor for each kinetics trace is then computed directly from the measured intensities in the two experimental setups: $\varsigma = I(0^-)/F_{eq}(T_{in})$. The time dependent change in fluorescence measured in the T-jump spectrometer in response to a T-jump, at $T_{fin} = T_{in} + \Delta T$, is written in terms of the time-dependent change in the population of the macrostates $P_i(t, T)$ from their initial population immediately after the T-jump, $\mathbf{P}(t = 0) = \mathbf{P_{eq}}(T_{in})$, to the final population at the end of the relaxation kinetics, $\mathbf{P}(t = \infty, T_{fin}) = \mathbf{P_{eq}}(T_{fin})$, and the eventual recovery of the this population back to $\mathbf{P_{eq}}(T_{in})$ as the T-jump itself recovers, as characterized by the $f_{rec}(t)$ function in Eq. S7:

$$I(t, T_{fin}) = \left[ \varsigma \sum_i F_i(T_{fin}) P_i(t, T_{fin}) - I(0^-) \right] f_{rec}(t) + I(0^-) \qquad (S19)$$

For a complete description of the equilibrium melting profiles and the relaxation kinetics traces in a self-consistent manner, the following variables are free parameters in our fitting algorithm: $\Delta H_i$ and $T_{mi}$ (to describe the free energy of each of the macrostates at all temperatures), $\Delta H_{i \to j}^{\ddagger}$ and $k_{i \to j}^0$ (to describe the forward rate constants between the macrostates at all temperatures, with the backward rates obtained from Eq. S17), $F_{io}$ and $m_i$ (to describe the fluorescence levels for each macrostate as a function of temperature), and $\Delta T$ and $\tau_{rec}$ for each kinetic trace. $I(0^-)$ is obtained from each experimental trace, by averaging the fluorescence intensities measured in each trace prior to the T-jump perturbation, and $T_{in}$ is obtained from measurements of the equilibrium sample temperature in the T-jump apparatus. The reference temperature for all equilibrium measurements was fixed at $T_0^{eq} = 20\ °C$. The reference temperature for the relaxation rates was fixed at $T^0 = 50\ °C$ for the pseudoknot to hairpin transition in VPK-2AP and at $T^0 = 85\ °C$ for the hairpin to single-stranded RNA transition, for VPK-2AP and HP1-2AP.

The global analysis approach was first applied to measurements on HP1-2AP, to simultaneously fit the fluorescence melting profile and kinetics traces measured at 4 different temperatures to two states (Figure 3a). The parameters describing the free energies, rate constants, and fluorescence levels for the hairpin and single-stranded conformations obtained from the fit to the HP1-2AP data were used to constrain the range of parameters of the hairpin (pkHP1) to single-stranded transition (U) in the kinetic scheme for the VPK pseudoknot (Figure 4a). In the case of VPK-2AP, the global analysis approach was applied to the fluorescence melting profile and kinetics traces measured at 9 different temperatures. The variable parameters and the experimental equilibrium and kinetics data were subjected to a simulated annealing fitting procedure (7–9). For HP1-2AP and VPK-2AP, 10,000 independent fitting procedures were performed.

The reported fitted parameters and their uncertainties were determined by two different methods. In the first method, the fitting parameters and errors were calculated as the weighted $(w = 1/\chi_v^2)$ average and standard deviation of the fitted parameters of all the fits between the range of 1-1.5 times the lowest $\chi_v^2$ obtained in independent fitting procedures. Here, $\chi_v^2$ is the reduced chisq, defined as $\chi^2$ (Eq. S6) divided by the number of data points minus the number of parameters. In the second method, the parameters and uncertainties were determined by the same procedure of error estimation used in van't Hoff analyses, as described in SI Methods 1.4. Errors determined using the second method are reported inside parentheses in SI Table S2. The values of the final temperatures reported in the kinetic traces shown in Figures 3 and 4 were obtained from the average T-jump obtained from the global fit for that set, for example from the average of the parameter $\Delta T$ for each of the 4 kinetics traces in Figure 3 and 9 kinetics traces in Figure 4.

The melting and kinetics fitting parameters allow the determination of the folding/unfolding rates at different temperatures, as specified in Eqs. S16 and S17. For unfolding rates, the uncertainties in the rates as a function of temperature were estimated in the following way:

$$\sigma_{k_{i \to j}}^- \left( \sigma_{k_{i \to j}^0} \right) = \left[ k_{i \to j}^0(T^0) - \sigma_{k_{i \to j}^0} \right] \exp\left[ -\frac{\Delta H_{i \to j}^\ddagger}{R}\left( \frac{1}{T} - \frac{1}{T^0} \right) \right] \qquad (S20)$$

$$\sigma_{k_{i \to j}}^+ \left( \sigma_{k_{i \to j}^0} \right) = \left[ k_{i \to j}^0(T^0) + \sigma_{k_{i \to j}^0} \right] \exp\left[ -\frac{\Delta H_{i \to j}^\ddagger}{R}\left( \frac{1}{T} - \frac{1}{T^0} \right) \right] \qquad (S21)$$

$$\sigma_{k_{i \to j}}^{-} \left( \sigma_{\Delta H_{i \to j}^{\ddagger}} \right) = k_{i \to j}^{0}(T^0) \exp \left[ -\frac{\left( \Delta H_{i \to j}^{\ddagger} - \sigma_{\Delta H_{i \to j}^{\ddagger}} \right)}{R} \left( \frac{1}{T} - \frac{1}{T^0} \right) \right] \quad (S22)$$

$$\sigma_{k_{i \to j}}^{+} \left( \sigma_{\Delta H_{i \to j}^{\ddagger}} \right) = k_{i \to j}^{0}(T^0) \exp \left[ -\frac{\left( \Delta H_{i \to j}^{\ddagger} + \sigma_{\Delta H_{i \to j}^{\ddagger}} \right)}{R} \left( \frac{1}{T} - \frac{1}{T^0} \right) \right] \quad (S23)$$

Each of the previous equations defines a curve imposing a lower (-) or upper limit (+) in the rates as determined by the errors of the fitting parameters $k_{i \to j}^{0}(T^0)$ and $\Delta H_{i \to j}^{\ddagger}$; the largest area enclosed by these curves is thus taken as a measure of the uncertainty in the calculation (see SI Figure S6). The uncertainties in the folding rates are calculated in a similar manner, this time depending on the uncertainties of $k_{i \to j}(T)$ (Eqs. S20–S23), $\Delta G_i$ and $\Delta G_j$ (whose uncertainties are determined by the fitting parameters $\Delta H_i$ and $T_{mi}$, Eq. S10). The unfolding/folding rate coefficients $k_{i \to j / j \to i}(T)$, computed at 37 °C are reported in SI Table S2.

**1.9 Simulation model and protocols**. We performed molecular dynamics simulations using a coarse-grained RNA model. The model and parameters were calibrated and described in details elsewhere (10). In the model, each nucleotide is represented by interaction sites (TIS) corresponding to a phosphate, ribose sugar, and base. Briefly, the effective potential energy of a given RNA conformation is U = $U_L$ + $U_{EV}$ + $U_{ST}$ + $U_{HB}$ + $U_{EL}$, where $U_L$ accounts for chain connectivity and angular rotation of the polynucleic acids, $U_{EV}$ accounts for excluded volume interactions, $U_{ST}$ and $U_{HB}$ are the base-stacking and hydrogen-bond interactions, respectively. The last term $U_{EL}$ is electrostatic interactions between phosphate groups, depending on a given salt concentration. Electrostatic interactions are taken into account using the Debye-Hückel theory to reproduce salt-dependent thermodynamics of RNA. The charges on the phosphates were determined using the counter ion condensation theory. The parameters in the TIS model were calibrated with thermodynamics of dinucleotides, several hairpins, and pseudoknots so that experimental data such as heat capacities are reproduced (10). The same model was also used to study another pseudoknot, BWYV PK, and verified to reproduce experimentally-obtained thermodynamics data (11). In particular for this study, we confirmed that the model gives excellent consistency with experimental heat capacity at both 50 mM and 1M monovalent salt (SI Figure S7). To obtain an equilibrium ensemble at 50 mM and 1 M, we used the temperature replica-exchange technique with low-friction Langevin dynamics to enhance the sampling efficiency (12, 13). For kinetic simulations, we first prepared the unfolded conformations by running simulations at 120 °C. Starting from the unfolded conformations, Brownian dynamics simulations under realistic water viscosity ($\eta = 10^{-3}$ Pa·s) were performed until each trajectory folded to the PK state at 37 °C. The same procedure was repeated at different salt concentrations of 50, 100, 200, 500, 800 mM and 1M. For each condition, we generated more than 200 folding trajectories. Each trajectory is categorized either pkHP1 or pkHP2 pathway according to which stem is folded earlier than the other. Each stem is deemed "folded" if all the base pairs are formed ($U_{HB} < k_B T$). For the purpose of converting simulation time into real time, we consider the characteristic time of the Brownian dynamics, $\tau = \frac{6\pi\eta a^3}{k_B T} \approx 300$ ps, where we use $a = 4$ Å as the typical length scale; because we integrate the motion by a small fraction of time, a single simulation step corresponds 3 ps (14, 15).

## SI REFERENCES

# SI TABLES

**Table S1.** Thermodynamic parameters of VPK and HP1 melting[a]

| Sample | $T_{m1}$ (°C) | | $\Delta H_1$ (kcal/mol) | | $T_{m2}$ (°C) | | $\Delta H_2$ (kcal/mol) | |
|---|---|---|---|---|---|---|---|---|
| | **Abs.** | **Fluor.** | **Abs.** | **Fluor.** | **Abs.** | **Fluor.** | **Abs.** | **Fluor.** |
| **VPK** | $47.3^{+2.6}_{-2.6}$ | - | $18.7^{+12.6}_{-2.9}$ | - | $85.0^{+1.6}_{-1.6}$ | - | $39.4^{+11.9}_{-10.2}$ | |
| | $(46.6^{+0.4}_{-0.4})$ | - | $(23.0^{+0.7}_{-0.7})$ | - | $(85.4^{+0.3}_{-0.3})$ | - | $(30.9^{+1.4}_{-1.4})$ | |
| **VPK-2AP** | $48.9^{+1.1}_{-1.0}$ | $49.5^{+1.0}_{-1.0}$ | $23.6^{+12.1}_{-1.7}$ | $61.6^{+10.5}_{-8.7}$ | $94.0^{+10.0}_{-7.6}$ | $84.3^{+2.0}_{-2.2}$ | $28.3^{+19.7}_{-2.9}$ | $43.5^{+10.7}_{-6.0}$ |
| | $(47.9^{+0.1}_{-0.1})$ | $(49.2^{+1.4}_{-0.6})$ | $(33.2^{+0.8}_{-0.8})$ | $(46.5^{+3.2}_{-3.2})$ | $(96.5^{+0.9}_{-0.3})$ | $(82.6^{+1.6}_{-0.4})$ | $(21.3^{+0.5}_{-0.5})$ | $(36.7^{+3.1}_{-2.3})$ |
| **VPK-pC** | $55.9^{+4.3}_{-2.1}$ | $41.5^{+0.7}_{-0.7}$ | $20.4^{+13.8}_{-9.6}$ | $33.1^{+1.7}_{-1.6}$ | $88.8^{+2.8}_{-1.7}$ | - | $54.3^{+10.0}_{-9.2}$ | - |
| | $(50.1^{+0.3}_{-0.3})$ | $(44.9^{+0.1}_{-0.1})$ | $(40.7^{+2.7}_{-2.6})$ | $(27.5^{+0.7}_{-0.6})$ | $(88.7^{+0.3}_{-0.2})$ | - | $(50.9^{+2.2}_{-2.1})$ | - |
| **HP1-2AP** | - | - | - | - | - | $82.9^{+0.2}_{-0.2}$ | - | $49.6^{+2.5}_{-2.5}$ |
| | - | - | - | - | - | $(85.4^{+0.2}_{-0.2})$ | - | $(35.7^{+0.4}_{-0.4})$ |

[a]Parameters and their errors were calculated as explained in SI Methods 1.3. The values in the top row are from fits to the absorbance or fluorescence melting profiles (*Method 1*); the values inside the parenthesis are from fits to the derivatives of the melting profiles (*Method 2*). The melting temperatures reported in the text are the average of the values from the two methods, rounded to the nearest integer.

**Table S2.** Folding/unfolding parameters obtained from kinetic modeling [a]

| Transition | $PK \rightarrow pkHP1$ [b] | $pkHP1 \rightarrow U$ [b] | $HP1 \rightarrow U$ [c,d] |
|---|---|---|---|
| $\Delta H$ (kcal/mol) | $31.4 \pm 5.0$ $(28.7^{+0.3}_{-0.3})$ | $42.9 \pm 1.8$ $(42.0^{+1.8}_{-1.8})$ | $40.6 \pm 3.2$ $(41.4^{+0.6}_{-0.6})$ |
| $T_m$ (°C) | $53.2 \pm 1.5$ $(54.4^{+0.1}_{-0.1})$ | $83.2 \pm 2.3$ $(81.9^{+0.1}_{-0.1})$ | $87.3 \pm 1.9$ $(86.9^{+0.1}_{-0.1})$ |
| $\Delta H^{\ddagger}_{i \rightarrow j}$ (kcal/mol) | $47.6 \pm 6.7$ $(44.8^{+3.8}_{-4.0})$ | $57.7 \pm 1.7$ $(53.5^{+21.0}_{-ND})$ | $47.6 \pm 4.6$ $(48.6^{+5.1}_{-6.5})$ |
| $k^{0}_{i \rightarrow j}$ (at $T^0$) (s$^{-1}$) | $10^{2.3 \pm 0.1}$ $(10^{2.3^{+0.1}_{-0.1}})$ $(T^0 = 50$ °C$)$ | $10^{4.6 \pm 0.2}$ $(10^{4.8^{+ND}_{-0.3}})$ $(T^0 = 85$ °C$)$ | $10^{4.5 \pm 0.1}$ $(10^{4.5^{+0.2}_{-0.2}})$ $(T^0 = 85$ °C$)$ |
| $\Delta H^{\ddagger}_{j \rightarrow i}$ (at 37 °C) (kcal/mol) | $16.2 \pm 11.7$ $(16.1^{+4.1}_{-4.3})$ | $14.8 \pm 3.5$ $(11.5^{+22.8}_{-ND})$ | $7.0 \pm 7.8$ $(7.2^{+5.7}_{-7.1})$ |
| $k_{i \rightarrow j}$ (at 37 °C) (s$^{-1}$) [e] | $9.3^{+5.2}_{-3.3}$ | $0.4^{+0.3}_{-0.1}$ | $1.0^{+0.2}_{-0.6}$ |
| $k_{j \rightarrow i}$ (at 37 °C) (s$^{-1}$) [e] | $118.6^{+65.3}_{-42.1}$ | $(3.6^{+2.1}_{-1.3})$ x $10^3$ | $(9.9^{+17.1}_{-6.3})$ x $10^3$ |

[a]The two values reported for each quantity come from different calculations of mean values and errors (see SI Methods 1.8). PK: pseudoknot; pkHP1: Hairpin 1 in the pseudoknot context; HP1: Hairpin 1; U: unfolded RNA. *ND*: Error could not be determined. [b]The values shown are the average of fitting parameters obtained from 263 fits out of 10,000 independent fitting procedures. [c] Parameters are for the 2-state model of HP1-2AP. [d]The values shown are the average of fitting parameters obtained from 550 fits out of 10,000 independent fitting procedures. [e]The unfolding/folding rate coefficients at 37 °C ($k_{i \rightarrow j / j \rightarrow i}$) were calculated from the rate coefficients at $T^0$ ($k^{0}_{i \rightarrow j / j \rightarrow i}$) and the corresponding energy barriers ($\Delta H^{\ddagger}_{i \rightarrow j / j \rightarrow i}$), with uncertainties calculated from Eqs. S20–S23.

**Table S3.** Folding times at 37 °C from simulations

| Pathway | Through HP1 | Through HP2 |
|---|---|---|
| $\tau$ $(U \rightarrow HP)^a$ (at 37 °C) | 70 μs | 50 μs |
| $\tau$ $(HP \rightarrow PK)^a$ (at 37 °C) | 1.51 ms | 110 μs |

[a]PK: pseudoknot; HP: Hairpin; U: unfolded RNA.

**Supplementary Figure Legends**

**Figure S1.** Thermodynamics of VPK melting from absorbance experiments. (a) The absorbance values of VPK, VPK-2AP and VPK-pC, measured at 260 nm, are plotted as a function of temperature. The absorbance profiles are normalized by dividing each curve by the absorbance measured at the lowest temperature, at 20 °C. (b–d) The first derivative of the absorbance with respect to temperature ($\delta A / \delta T$) is plotted as a function of temperature for (b) VPK, (c) VPK-2AP and (d) VPK-pC. Prior to calculating the derivative, the absorbance data was smoothed using a moving average with the smooth built-in function in MATLAB. All experiments were done in 10 mM MOPS buffer pH 7.0, at 50 mM KCl. The data (symbols) are the averages of 2–3 independent sets of measurements. The error bars represent the standard deviation of the mean; some error bars are smaller than the symbols. For clarity, only every other data point is presented. The continuous lines are fits to two sequential van't Hoff transitions, with fitting parameters summarized in SI Table S1. Unlabeled VPK exhibits two melting transitions at ~47 °C and ~85 °C; the corresponding transitions for VKP-2AP occur at ~49 °C and ~95 °C, and for VPK-pC at ~53 °C and ~ 89 °C.

**Figure S2.** Estimation of uncertainty in fitting parameters. The plot illustrates how the variation in $\chi^2$ as a function of a parameter in a fit is used to estimate the uncertainty in that parameter. The $\chi^2$ values obtained from a fit to the VPK-2AP fluorescence melting profile, for the data shown in Fig. 3b, are plotted as a function of the melting temperature $T_{m1}$; $\chi^2$ values were found as described in SI Methods 1.4. The red line indicates the value at which $\Delta \chi^2 = 1$ , which was used as a cutoff for the uncertainties in that parameter, as indicated by the dashed vertical lines. In this example, the value of the parameter and its uncertainty thus estimated is $49.5^{+1.0}_{-1.0}$.

**Figure S3.** Equilibrium fluorescence measurement on the control sample pC-ref (5'-GC[pC]CA-3'). The fluorescence of pC-ref, with excitation at 350 nm, is plotted as a function of temperature. The data (symbols) are the average of two independent sets of measurements. The error bars represent the standard deviation of the mean. For clarity, only every other data point is presented.

**Figure S4.** Thermodynamics of HP1-2AP melting from fluorescence experiments. (a) 2AP fluorescence of HP1-2AP and (b) the corresponding first derivative $\delta F / \delta T$ is plotted as a function of temperature. All experiments were done in 10 mM MOPS buffer pH 7.0, at 50 mM KCl. The data (symbols) are the averages of two independent sets of measurements. The error bars represent the standard deviation of the mean; some error bars are smaller than the symbols. For clarity, only every other data point is presented. The continuous black lines are fits to a single van't Hoff transition occurring at ~84 °C; the green line corresponds to the lower baseline of the fit. All fitting parameters are summarized in SI Table S1.

**Figure S5.** Equilibrium fluorescence and T-jump measurements on the control sample 2AP-ref (5'-CC[2AP]CU-3'). (a) The fluorescence of 2AP-ref, with excitation at 310 nm, is plotted as a function of temperature. The data (symbols) are the average of two independent sets of measurements. The error bars represent the standard deviation of the mean. For clarity, only every other data point is presented. The continuous black line is a linear fit to the data. (b) The fluorescence intensities measured on a 2AP-ref sample in the T-jump spectrometer are plotted as a function of time. The negative values on the $x$-axis indicate measurements prior to the arrival of the IR pulse that induces the T-jump in the sample. Immediately after the T-jump, the fluorescence in the sample drops, as a result of the change in the quantum yield of 2AP. The fluorescence then decays back to the pre-T-jump levels. The continuous line is a fit to the data,

using the T-jump recovery function described by Eq. S8, with a characteristic recovery time constant $\tau_{rec} = 143$ ms.

**Figure S6.** Folding/unfolding rates of VPK-2AP. The rate constants for transitions between the three states depicted in Figure 4a, calculated using the parameters obtained from the global fit to the data in Figure 4, are plotted as a function of temperature. The shaded areas depict the uncertainties in these rate constants, which are computed as described in the text (Eqs. S20-S23).

**Figure S7.** Thermodynamics of VPK obtained by equilibrium simulations at 50 mM and 1 M salt concentrations. (a) Heat capacity of VPK at 50 mM. The two observed transitions are at 47°C and 84°C. (b) Same as (a) but at 1M salt concentration. The two transitions in (b) occur at 70°C and 88°C.

**Figure S8.** Thermodynamics of VPK unfolding from coarse-grained simulations. The equilibrium populations of the folded and unfolded conformations of VPK, described in terms of a 4-state model (Figure 6a), are plotted as a function of temperature, from simulations at (a) 50 mM and (b) 1 M monovalent salt. Populations are shown for the fully folded VPK (pink), partially folded hairpins pkHP1 and pkHP2 (green and orange, respectively), and fully unfolded U (blue).

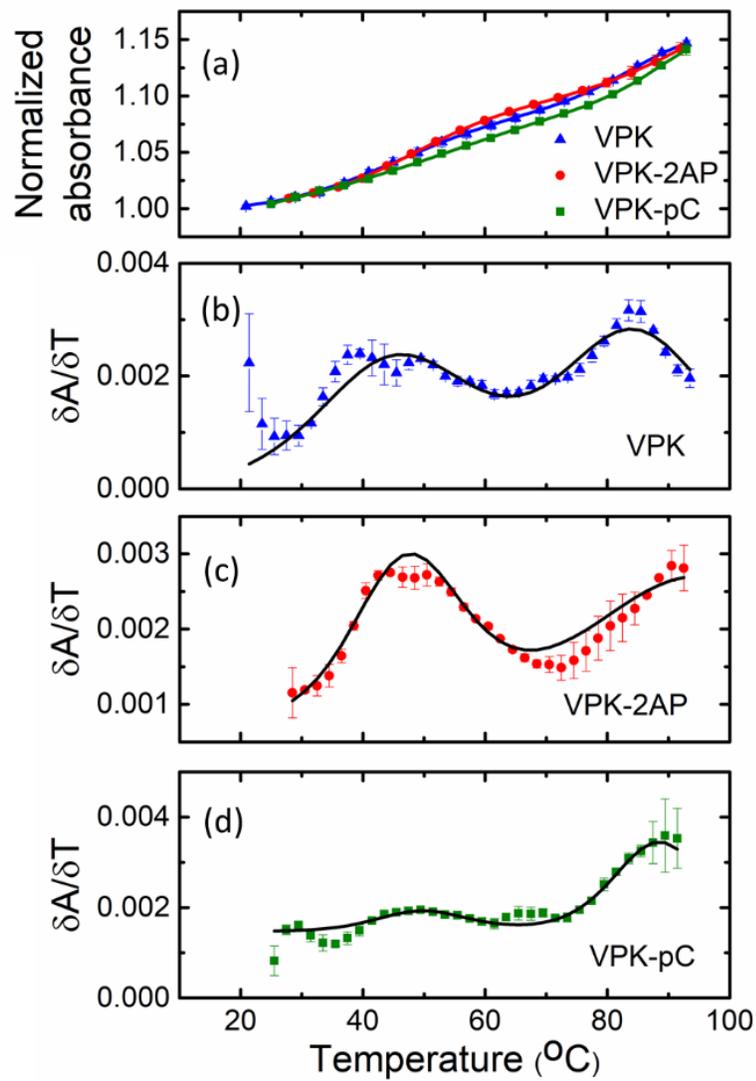

**Roca et al.
Supplementary Figure S1**

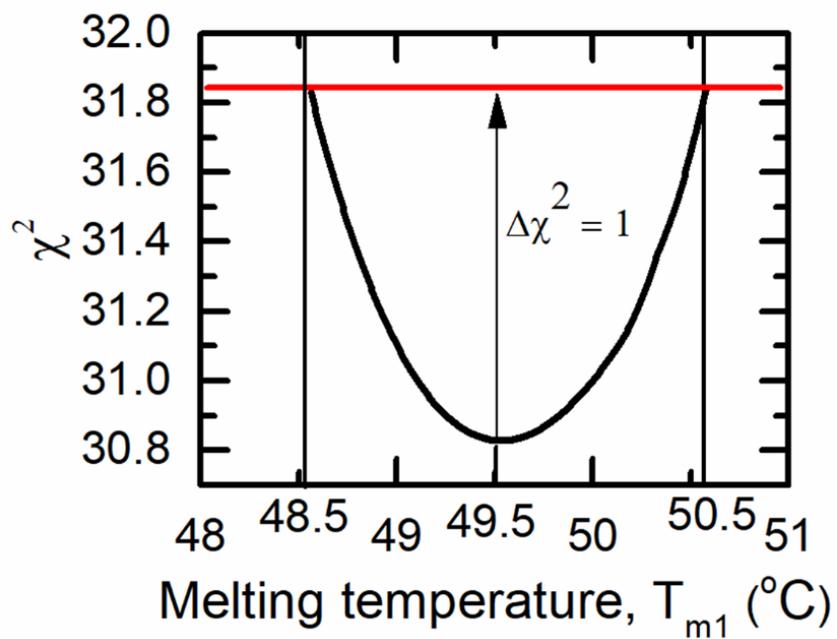

**Roca et al.**
**Supplementary Figure S2**

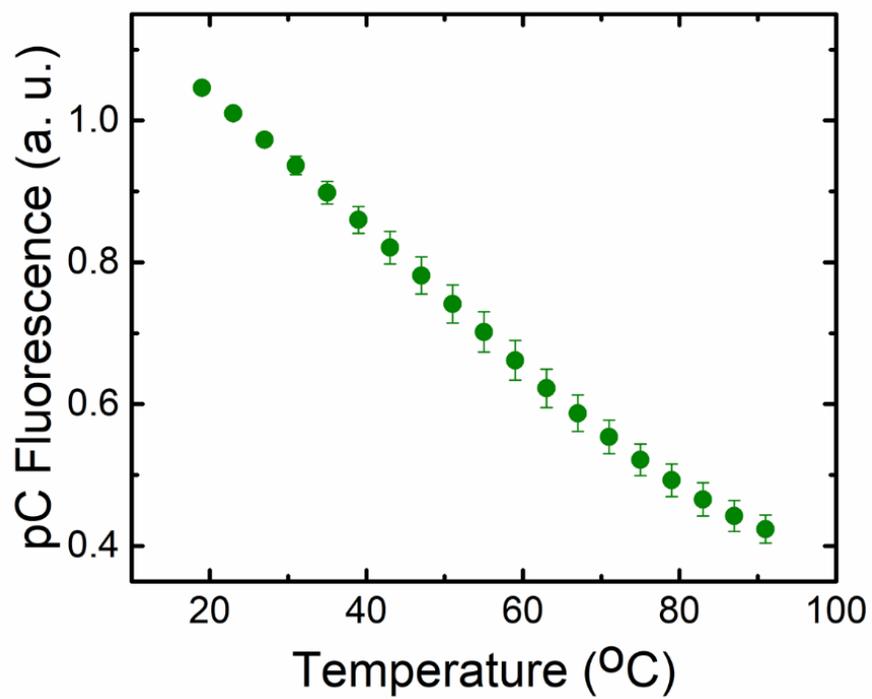

**Roca et al.**
**Supplementary Figure S3**

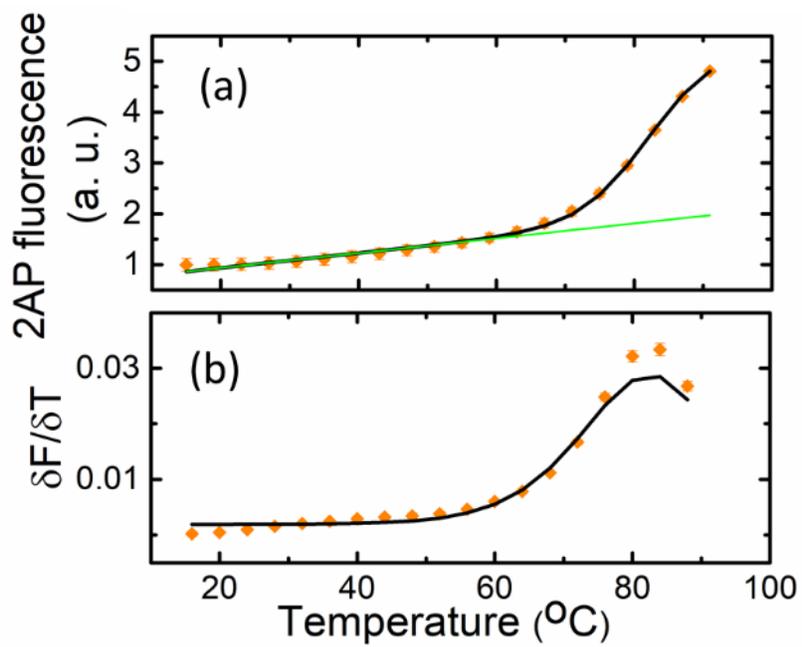

**Roca et al.**
**Supplementary Figure S4**

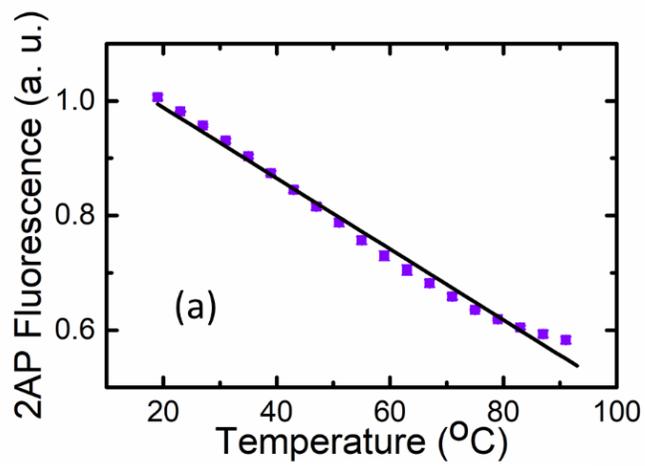
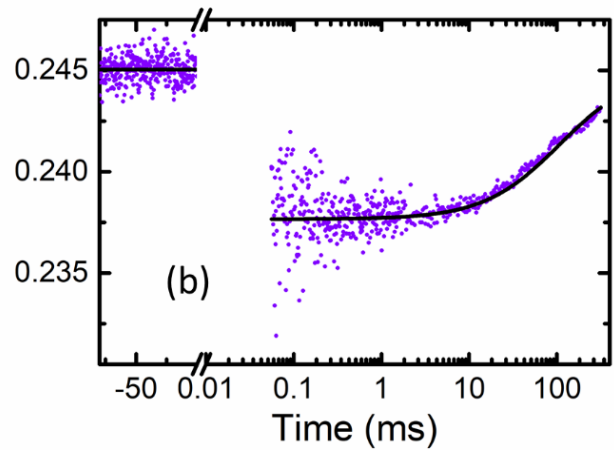

**Roca et al.**
**Supplementary Figure S5**

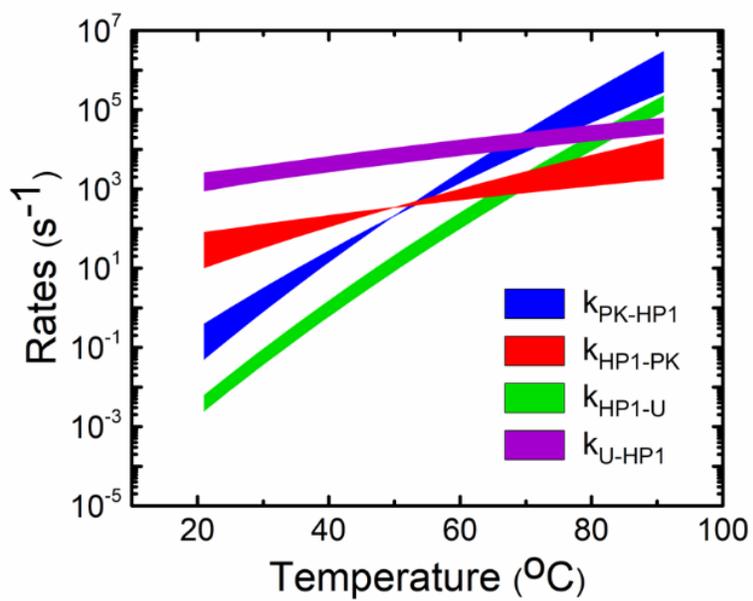

**Roca et al.**
**Supplementary Figure S6**

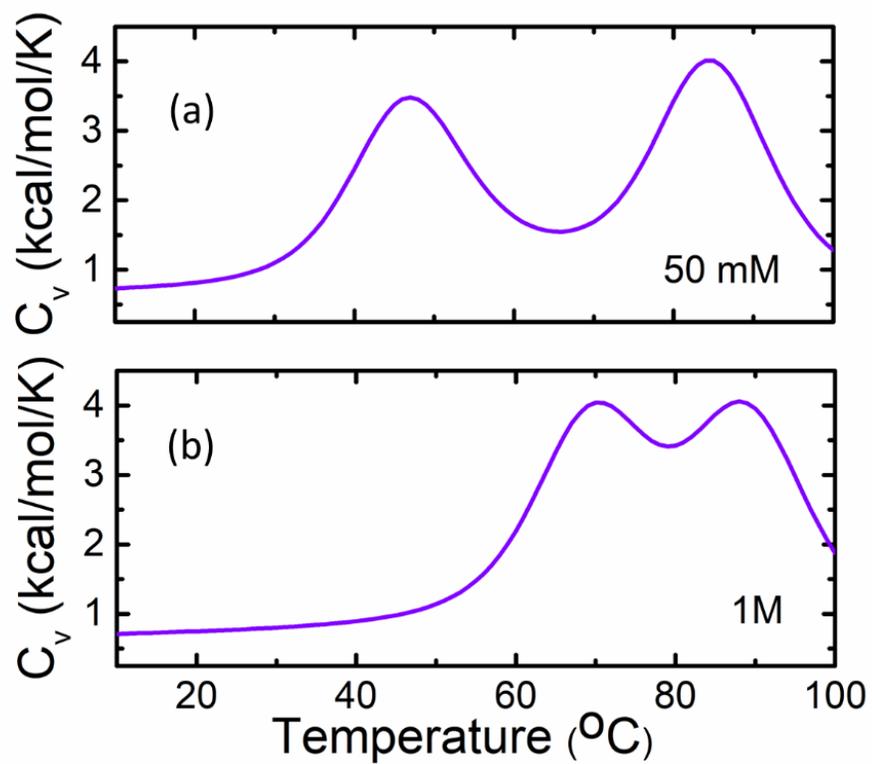

Roca et al.
Supplementary Figure S7

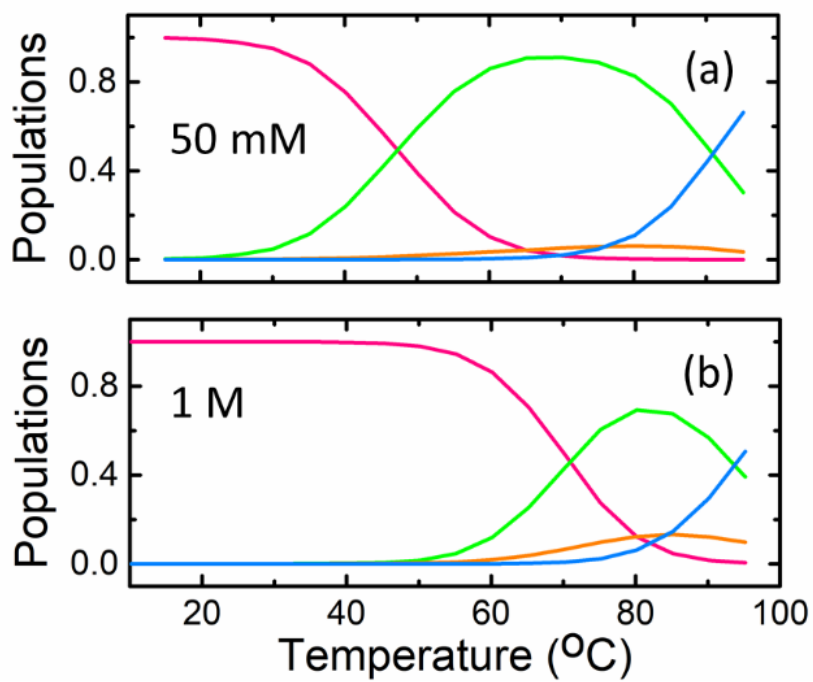

Roca et al.
Supplementary Figure S8